\documentclass[epj]{svjour}
\usepackage{graphicx,amsmath,amssymb}
\newcommand{\eg}{\textit{e.g.}}

\newcommand{\dd}{\mathrm{d}}

\begin{document}
\title{Forward and midrapidity charmonium production at RHIC}
\author{Xingbo Zhao\inst{1} \and Ralf Rapp\inst{1}
}                     
%
%
\institute{Cyclotron Institute and Physics Department, Texas A\&M University, College Station, TX 77843-3366, USA}
\date{Received: date / Revised version: date}
%
\abstract{ $J/\psi$ production at forward and midrapidity at the
  Relativistic Heavy-Ion Collider (RHIC) is calculated within a
  previously constructed rate-equation approach accounting for both
  direct production and regeneration from $c$ and $\bar c$. The
  results are compared to experimental data. The observed stronger
  suppression at forward rapidity can be qualitatively explained by a
  smaller statistical regeneration component together with stronger
  cold nuclear matter induced suppression compared to midrapidity. The
  $\chi_c$ over $J/\psi$ ratio and $\psi'$ over $J/\psi$ ratio are
  also calculated.
} 
\maketitle
\section{Introduction}
\label{intro}
The suppression of $J/\psi$ mesons in ultrarelativistic heavy-ion
collisions, as a result of color Debye screening, has been suggested
as a signature~\cite{Matsui:1986dk} of Quark-Gluon Plasma (QGP)
formation a long time ago and is indeed observed in both $Pb$-$Pb$
collisions at the CERN Super Proton Synchrotron
(SPS)~\cite{Ramello:2003ig} and in $Au$-$Au$ collisions at BNL
Relativistic Heavy-Ion Collider (RHIC)~\cite{Adare:2006ns}. However,
it is puzzling that the observed suppression is very comparable at SPS
and RHIC energies, since the energy density of the medium at RHIC is
expected to be much higher than at SPS. The statistical
model~\cite{BraunMunzinger:2000ep}, or kinetic
approaches~\cite{Grandchamp:2003uw,Zhao:2007hh}, explain this puzzle
by considering the regeneration of charmonium from $c$ and $\bar c$
quarks: at RHIC energies more $c$ and $\bar c$ quarks (relative to
charmonium) are produced than at SPS energies and therefore the
regeneration of charmonium at RHIC largely compensates the expected
stronger suppression (which was, in fact, predicted in
Ref.~\cite{Grandchamp:2001pf}). Recent RHIC data~\cite{Adare:2006ns}
suggest another puzzle: charmonium suppression observed at forward
rapidity ($|y|\in$[1.2,2.2]) is stronger than at midrapidity
($|y|<$0.35), despite the energy density of the medium at forward
rapidity being presumably smaller which should lead to weaker
suppression.

$J/\psi$ production at forward rapidity has been investigated by
several theoretical models. In the kinetic recombination
model~\cite{Thews:2005vj} the rapidity and transverse momentum
distributions of $J/\psi$ are obtained through solving a rate equation
with the gain term accounting for the continuous formation process of
the $J/\psi$'s from $c$ and $\bar c$ quarks throughout the QGP. The
``input'' charm-quark spectra are obtained from either perturbative
QCD (pQCD) calculations or thermal distributions and the inelastic
reaction employed is the traditional gluo-dissociation
process~\cite{Peskin:1979va}. A narrowing of the $J/\psi$ rapidity
distribution is predicted in $A$-$A$ relative to $p$-$p$
collisions. In the statistical hadronization model
(SHM)~\cite{Andronic:2006ky} all charmonia are produced from
coalescence of $c$ and $\bar c$ quarks at the hadronization
transition. The relative abundance of open and hidden charm states is
determined by their mass and spin-isospin degeneracy, based on the
assumption of thermal equilibrium at the hadronization temperature,
$T_c$. The resulting $J/\psi$ rapidity distributions from the SHM are
also narrower in $A$-$A$ than in $p$-$p$ collisions, but significantly
wider than those from the kinetic recombination approach, mostly due
to broader input distributions of charm-quark cross section. In the
comovers interaction model (CIM)~\cite{Armesto:1997sa} the
primordially produced $J/\psi$'s are subject to a series of cold
nuclear matter induced (initial state) effects and subsequently
destroyed by ``comovers'' in the medium with an effective dissociation
cross section; $\sigma_{Diss}\sim$0.65-1~mb. The CIM has recently been
augmented by including charm quark coalescence into
$J/\psi$~\cite{Capella:2007jv}. The rapidity dependence of
experimental data is reproduced both for $Au$-$Au$ and $Cu$-$Cu$
collisions at RHIC, mostly due to initial-state effects in the
incoming nuclei.

In the present work we apply a rate-equation
approach \cite{Grandchamp:2003uw,Zhao:2007hh} to 200 AGeV $Au$-$Au$
collisions at RHIC, to calculate $J/\psi$ production at forward
rapidity and compare to previously obtained midrapidity
results~\cite{Zhao:2007hh}. The approach assumes a thermalized medium
(QGP and subsequent had\-ronic gas (HG)) in which anomalous suppression
and regeneration from coalescence of $c$-$\bar c$ quarks and $D$-$\bar
D$ mesons occur. In Section~\ref{initial} we evaluate cold nuclear
matter effects which affect the charmonium abundances before the
medium thermalizes. We then recall basic ingredients of our approach
to evaluate $J/\psi$ suppression and regeneration in QGP and HG phase
in Section~\ref{supp}. We discuss the numerical results for the
centrality dependence of the inclusive $J/\psi$ yield and $\langle
p_t^2\rangle$ as well as $\chi_c$ to $J/\psi$ and $\psi'$ to $J/\psi$
ratios in Section~\ref{num} and the $J/\psi$ transverse momentum
spectra in Section~\ref{pt}. Conclusions are given in
Section~\ref{concl}.

\section{Cold nuclear matter effects}
\label{initial}
A ``pre-charmonium'' $c\bar c$ pair produced in an initial hard
nucleon-nucleon ($N$-$N$) collision first travels through the incident
(cold) nuclei before becoming a fully formed charmonium ($\Psi=J/\psi,
\chi_c, \psi'$). The modification of initial parton distribution
functions, affecting the hard production of pre-charmonium, and the
interactions between pre-$\Psi$ and cold nuclei are referred to as
cold nuclear matter (CNM) effects. The latter provide a baseline for
identifying the anomalous suppression and/or enhancement of charmonia
which is hoped to give insights about the subsequent hot medium. An
accurate estimate of CNM effects is therefore mandatory. In the
present work we assume CNM effects to be the only relevant ones to
charmonium production before the medium thermalizes and investigate
them within the following two schematic baseline scenarios: (1)
Nuclear absorption + Cronin effect. (2) Shadowing + Nuclear absorption
+ Cronin effect.

Let us start with scenario 1: The initial dissociation by primordial
nucleons passing by is described by the standard Glauber model
resulting in a spatial charmonium distribution at the thermalization
time $\tau_0$,
\begin{eqnarray}
  f_\Psi(\vec{x}_t,\tau_0)=\sigma^\Psi_{pp}\int d^2s\ dz\ dz^{\prime}\rho_A(\vec{s},z)\
  \rho_B(\vec{x}_t-\vec{s},z^{\prime})\nonumber\\ \times\exp\left\{-\int^\infty_z 
    dz_A\rho_A(\vec{s},z_A)\sigma_{nuc}\right\}\nonumber\\
  \times\exp\left\{-\int^\infty_{z^\prime} 
    dz_B\rho_B(\vec{x}_t-\vec{s},z_B)\sigma_{nuc}\right\}\ ,
\label{fx_glauber}
\end{eqnarray}
where $\rho_{A,B}$ are Woods-Saxon profiles~\cite{De Jager:1974dg} of
nuclei $A$ and $B$. The nuclear absorption cross section,
$\sigma_{nuc}$=1.5~mb ($\sigma_{nuc}$=2.7~mb for $\psi'$), serves as a parameter to regulate the strength
of dissociation due to all CNM effects combined, estimated from
midrapidity $d$-$Au$ collisions~\cite{Adare:2007gn}.  Shadowing
effects are effectively parameterized into $\sigma_{nuc}$, which we
furthermore assume to be rapidity-independent. Such a scenario may be
justified if shadowing in one $Au$ nucleus (forward rapidity in
$d$-$Au$) is roughly compensated by anti-shadowing in the other $Au$
nucleus (backward rapidity in $d$-$Au$). In addition to absorption,
transverse momentum ($p_t$) spectra of (pre)$\Psi$'s are broadened
compared to $p_t$ spectra in $p$-$p$ collisions due to the Cronin
effect. We assume the physical mechanism of the Cronin effect to be
the rescattering of gluons in the CNM before they fuse into
charmonium.  Therefore, the increase of $\langle p^2_t\rangle$ is
proportional to the mean length, $\langle l\rangle$, the gluon travels
in CNM assuming a random walk treatment of the gluon rescattering,
resulting in $\Delta \langle p_t^2\rangle=\langle
p_t^2\rangle_{AA}-\langle p_t^2\rangle_{pp}=a_{gN}\langle
l\rangle$. The proportionality coefficient $a_{gN}$ can in principle
be estimated from $d$-$Au$ data: here we adopt $a_{gN}$=0.1~GeV$^2$/fm
which is compatible with SPS NA50 data~\cite{Topilskaya:2003iy} and
current RHIC data~\cite{Adare:2007gn}. At forward rapidity the
$d$-$Au$ data show a $p_t$ broadening of $\Delta \langle
p_t^2\rangle$=$\langle p_t^2\rangle_{dAu}-\langle
p_t^2\rangle_{pp}$=0.8$\pm$0.4~GeV$^2$~\cite{Adare:2007gn} (we
combined the uncertainty for $\langle p^2_t\rangle$ of $d$-$Au$ and
$p$-$p$ in quadrature). With a ``path length'' $\langle l\rangle$=4~fm
for $d$-$Au$ collisions~\cite{Topilskaya:2003iy} we estimate an
uncertainty of $a_{gN}$=0.1$\sim$0.3~GeV$^2$/fm in our numerical
calculations and then we perform a Gaussian smearing over the input
power-law $p_t$ spectra from $p$-$p$ data with $\Delta \langle
p^2_t\rangle$=$a_{gN}\langle l\rangle$.

Concerning scenario 2 we follow the treatment of
Ref. \cite{Capella:2007jv} and separately treat nuclear absorption,
shadowing and Cronin effect. The total suppression by CNM effects
(nuclear absorption + shadowing) in this scenario is comparable to the
scenario 1 at midrapidity but is stronger at forward rapidity due to
stronger shadowing at forward rapidity. Concerning $p_t$ distributions
and Cronin effect our treatment in this scenario is identical to
scenario~1.

\section{Kinetic charmonium production approach}
\label{supp}
We assume that the medium formed at RHIC reaches thermal equilibrium
at about $\tau_0$=1/3~fm/$c$ and use a thermal rate equation thereafter to describe
the subsequent evolution of $\Psi$'s according to,
\begin{equation}
\label{rate-eq}
\frac{\dd N_{\Psi}}{\dd
  \tau}=-\Gamma_{\Psi} \ (N_{\Psi}-N_{\Psi}^{\text {eq}})
\ ,
\end{equation}
(as usual, we account the feeddown from $\chi_c$'s and $\psi'$'s to
$J/\psi$'s; formation time effects, as considered in
Ref.~\cite{Zhao:2008vu}, are not included in the present analysis). The initial condition follows from CNM
effects discussed in the previous section. The loss term,
$-\Gamma_{\Psi}N_{\Psi}$, accounts for the dissociation of the
primordially produced charmonium and the gain term,
$\Gamma_{\Psi}N_{\Psi}^{\text{eq}}$, for the regeneration of charmonia
via coalescence of $c$ and $\bar c$ quarks. $\Gamma_{\Psi}$ is the
in-medium charmonia dissociation rate and $N_{\Psi}^{\text {eq}}$ is
the equilibrium limit of the charmonium abundances. The rate equation
treats the coalescence of $c$ and $\bar c$ quarks as a continuous
process over the course of medium evolution in line with lattice
QCD~\cite{Asakawa:2003re} and potential-model
results~\cite{Cabrera:2006wh} which suggest the $J/\psi$ to survive at
temperatures well above the critical temperature $T_c$.

Let us first address the dissociation of primordially produced
charmonia (direct component), which is identified as the solution of
the homogeneous rate equation with only the loss term included. In our
previously constructed kinetic approach a Boltzmann transport equation
is employed to describe the evolution of the charmonium phase space
distribution functions $f_\Psi(\vec{x},\vec{p},\tau)$ in a thermalized
medium with initial distribution from Eq.~(\ref{fx_glauber}),
\begin{equation}
p^{\mu}\partial_{\mu}f_\Psi(\vec{x},\vec{p},\tau)
=-E_\Psi \ \Gamma_\Psi(\vec{x},\vec{p},\tau) \ 
f_\Psi(\vec{x},\vec{p},\tau) \ .
\label{boltz}
\end{equation}
We neglect elastic charmonium rescattering and only account for
parton(hadron)-induced inelastic scattering. The momentum-dependent
dissociation rate $\Gamma_{\Psi}$ is calculated via a convolution of
the charmonium dissociation cross section, $\sigma^{diss}$, with a
thermal distribution $f(\omega;T)$ of particles from the heat bath
(QGP or HG) and their relative velocity $v_{rel}$, with
charmonia~\cite{Zhao:2007hh},
\begin{equation}
\label{gamma}
\Gamma_\Psi(\vec{x},\ \vec{p},\ \tau) 
= \sum_{i} \int \frac{d^3k}{(2\pi)^3} \ f^i(\omega_k;T(\tau)) 
 \  \sigma^{diss}_{\Psi i} \ v_{rel}\ .
\end{equation}
In QGP we include the in-medium reduction of charmonium binding energy
due to color Debye screening. Under these circumstances the
gluo-dissociation process $\Psi+g\to c+\bar c$~\cite{Peskin:1979va} is
ineffective to destroy a charmonium~\cite{Grandchamp:2001pf} and we
instead employ a quasifree dissociation process: $i + \Psi \to
i+c+\bar c$ ($i$=$g$,$q$,$\bar q$)~\cite{Grandchamp:2001pf}. The
strong coupling constant $\alpha_s$ in the quasifree cross section is
one of the two main parameters in our approach and is adjusted to
reproduce the $J/\psi$ yield in central $Pb$-$Pb$ collisions at SPS
(resulting in $\alpha_s$=0.24, cf. the upper panel of
Fig.~\ref{raa_centra_mid}). The comparison between the charmonium
3-momentum dependence of gluo-dissociation (with vacuum binding
energy) and quasifree rates (with in-medium binding energy) can be
found in Ref.~\cite{Zhao:2007hh}: these two dissociation rates are
comparable for low-momentum $J/\psi$ but at high momentum the
gluo-dissociation rate is significantly smaller than the quasifree
one; for $\chi_c$ the gluo-dissociation rate is much larger than
quasifree rate. It is noteworthy that the so-called ``leakage
effect''~\cite{Karsch:1987,Blaizot:1988ec,Huefner:2002tt} is accounted
for by the Boltzmann transport equation approach: high $p_t$ charmonia
which travel outside the fireball are not subject to suppression
anymore rendering less suppression.

\begin{figure}[!t]
\centering
\includegraphics[width=0.40\textwidth]{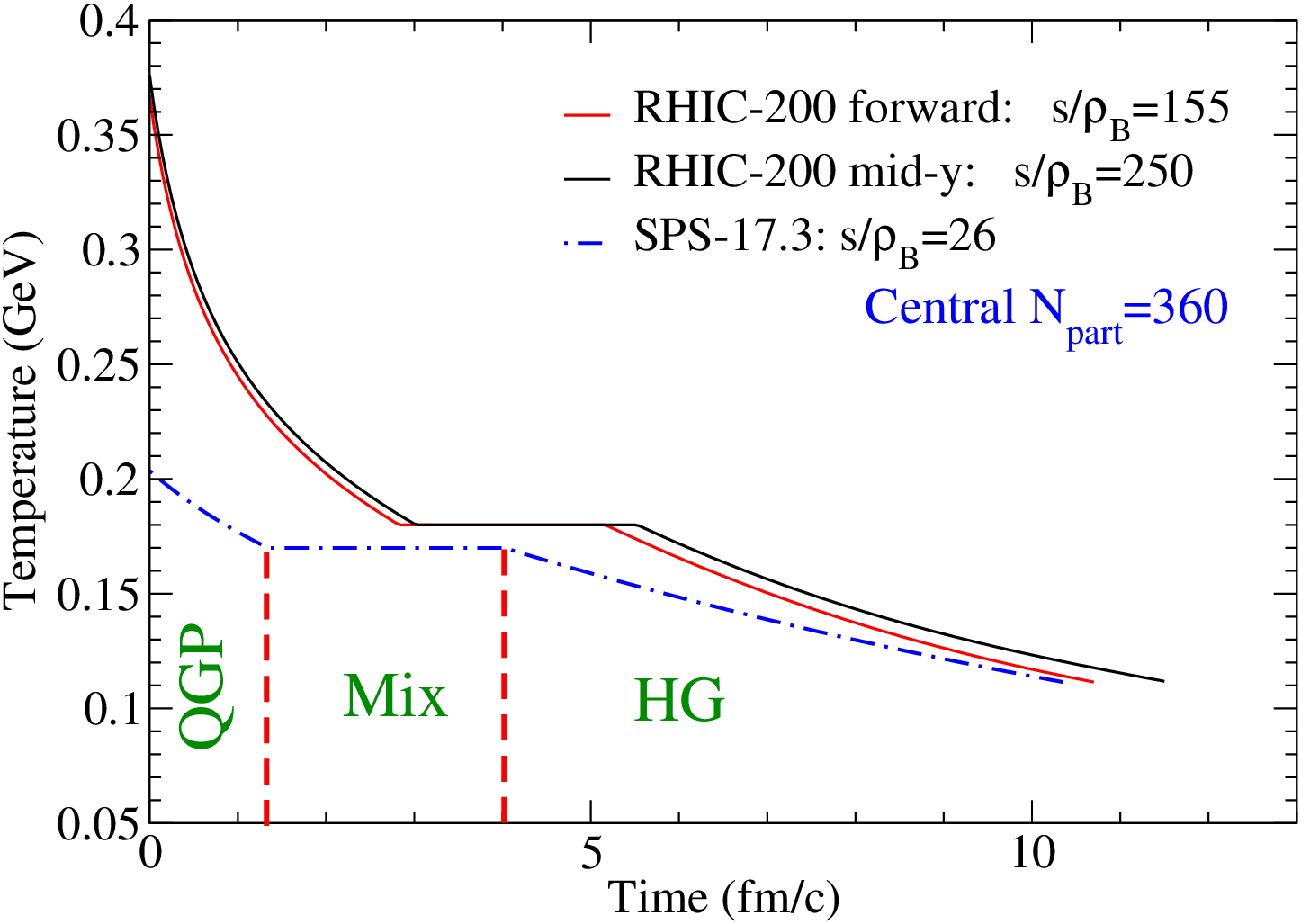}
\caption{(Color online) Temperature profiles at RHIC 200 AGeV
  (solid line: midrapidity; dashed line: forward rapidity)
  and at SPS 17.3 AGeV (dot-dashed line) for central collisions
  ($N_{part}$=360) of heavy nuclei.}
\label{fig_fb}
\end{figure}

To solve the transport equation~(\ref{boltz}) we need to know the
medium temperature evolution, $T(\tau)$, which is determined in the
following way: We assume the total entropy to be produced at the
thermalization time ($\tau_0$=1/3~fm/$c$ at RHIC). The subsequent
expansion is approximated by a cylindrically and isentropically
expanding fireball,
\begin{equation}
\label{Vfb}
V_{FB}(\tau)=(z_0+v_z\tau) \ \pi \ 
(r_0+\frac{1}{2}a_\perp \tau^2)^2\ .
\end{equation}
A freeze-out temperature of $T_{fo}\simeq120$~MeV terminates the
evolution and results in $\tau_{fo}=10-15$~fm/$c$. The fireball covers
1.8 rapidity units and the initial transverse radius $r_0$ represents
the initial transverse overlap of the two colliding $Au$ nuclei at a
given impact parameter $b$. The expansion parameters
${v_z,a_z,a_{\perp}}$ capture the main aspects of hydrodynamic
calculations to reproduce observed flow velocities. For simplicity we
assume that the expansion parameters are the same for midrapidity
($|y|<$0.35) and forward rapidity ($|y|\in$[1.2,2.2]). The total
entropy is, however, different, being inferred from number of the
charged-particle rapidity density, for which we use BRAHMS
data~\cite{Arsene:2004fa} $\dd N_{ch}/dy$=660 at midrapidity and $\dd
N_{ch}/dy$=615 at forward rapidity (for 0-5\% centrality);
PHOBOS~\cite{Back:2004je} and PHENIX~\cite{Adcox:2004mh} give
consistent results with slightly larger charged particle
multiplicities at both midrapidity and forward rapidity. For the
equation of state (EoS) we use an ideal gas of massive quarks and
gluons for $T>T_c$ and a resonance gas equation of state for $T<T_c$
including the 37 lowest lying mesons and 37 lowest lying baryons. The
critical temperature, $T_c$=180~MeV, at RHIC 200 AGeV is in line with
thermal-model fits of particle ratios and the current predictions of
lattice QCD~\cite{Cheng:2006qk}. Combining entropy conservation and
EoS the temperature evolution profile follows as a function of time
$\tau$ as displayed in Fig.~\ref{fig_fb} for central $Au$-$Au$
collisions at both midrapidity and forward rapidity, and compared to
SPS $Pb$-$Pb$ $\sqrt s$=17.3 AGeV collisions. We can see that the
lifetime of QGP at forward rapidity is a slightly shorter than at
midrapidity due to less produced particles and smaller energy
densities.

Now we turn to the regeneration component which is identified as the
difference between the solution of the full rate equation
(\ref{rate-eq}) and the equation without the gain term,
$\Gamma_{\Psi}N_{\Psi}^{\text{eq}}$ on r.h.s. The latter is dictated
by detailed balance with $\Gamma_{\Psi}$ the same (dissociation) rate
for the direct component. For the charmonium abundances in equilibrium
limit, $N_{\Psi}^{\text {eq}}$, we employ the statistical
model~\cite{Andronic:2006ky} with $N_{\Psi}^{\text
  {eq}}$=$N_{\Psi}^{\text {stat}}\mathcal R$, where $\mathcal R$ is a
correction coefficient which will be detailed later, and
$N_{\Psi}^{\text {stat}}$ is the abundance from the statistical model.
This assumes that all the $c$ and $\bar c$ quarks are exclusively
produced primordially, and populate open and hidden charm states
according to relative chemical equilibrium. The equilibrium abundance
of regenerated charmonia is given by
\begin{equation}
 N_{\Psi}^{stat}=\gamma_c^2Vn_{\Psi}
\label{eq:Jpsi_coal}
\end{equation}
with $V$: the volume of the fireball, $n_{\Psi}$: chemical
equilibrium density of charmonium $\Psi$, and $\gamma_c$: charm
quark fugacity. The latter is determined by the canonical charm conservation
equation,
\begin{equation}
N_{c\bar c}=\frac{1}{2} \gamma_c Vn_{op}
\frac{I_1(\gamma_c Vn_{op})}{I_0(\gamma_c Vn_{op})}+
\gamma_c^2 Vn_{hid}
\label{eq:Ncc}
\end{equation}
with $N_{c\bar c}$: total number charm quark pairs produced, $n_{op}$:
density of all open charm states, $n_{hid}$: density of all hidden
charm states. In the present work we use the charm quark cross section
$\dd \sigma_{\bar cc}/\dd y (y=0)$=95~$\mu$b in line with recent
PHENIX measurements~\cite{Adare:2006hc} (with the assumption that 30\%
of the total produced charm are uniformly distributed over 1.8
rapidity units which is the rapidity coverage of our fireball). For
the rapidity dependence we take guidance from perturbative QCD
calculation~\cite{Cacciari:2005rk}, leading to $\dd \sigma_{\bar
  cc}/\dd y (y$=$1.7)$=60~$\mu$b. Currently the uncertainties of charm
production in both theoretical predictions and experimental
measurements are rather large. The modified Bessel function factor
$\frac{I_1(\gamma_c Vn_{op})}{I_0(\gamma_c Vn_{op})}$ on the r.h.s
accounts for exact $c\bar c$ conservation in the canonical ensemble
relative to grand-canonical
ensemble~\cite{Cleymans:1990mn,Gorenstein:2000ck}.

The abundance of charmonia, $N_{\Psi}^{stat}$, from the statistical
model is subject to two additional corrections to obtain
$N_{\Psi}^{eq}$ in the rate equation Eq.(\ref{rate-eq}). The first
correction simulates incomplete charm quark thermalization in medium:
It is natural to expect that the coalescence rate from non-thermalized
$c$ and $\bar c$ quarks is smaller than for fully thermalized charm
quarks~\cite{Greco:2003vf,Yan:2006ve}. We implement this correction by
multiplying the charmonium abundances from the statistical model in a schematic way by a
factor $\mathcal R=1-\exp (-\tau/\tau_{eq}^c)$, where
$\tau_{eq}^c$ is the thermal relaxation time of charm quarks which is
the second main parameter in our approach. The second correction is a
``correlation volume'' $V_{corr}$ for the coalescing charm
quarks~\cite{Grandchamp:2003uw}. The correlation volume increases
$\Psi$ production because the $c\bar c$ pairs do not have time to
populate the entire fireball volume. We implement this correction by
replacing the fireball volume $V$ in the argument of the Bessel
function by a correlation volume $V_{corr}$ in Eq.(\ref{eq:Ncc}). The
``correlation volume'' is estimated as the volume explored by a
receding $c\bar c$ pair: $V_0(\tau)$=$4\pi(r_0 + \langle v_c\rangle
\tau)^3/3$, where $r_0$$\simeq$1.2~fm represents a minimal radius
characterizing the range of strong interactions, and $\langle
v_c\rangle$ denotes the average relative speed of the produced $c$ and
$\bar c$ quarks which we vary between 0.5$c$ and 0.8$c$. Within this
setup we adjust our second parameter $\tau_{eq}^c$ to the inclusive
$J/\psi$ yield in central $Au$-$Au$ at RHIC at midrapidity, resulting
in $\tau_{eq}^c$=4-7~fm/$c$, corresponding to $\langle
v_c\rangle$=0.8$c$ and 0.5$c$, respectively.

The transverse momentum distributions of regenerated charmonia are
approximated by local thermal equilibrium and boosted by the
transverse flow of the medium, corresponding to a blastwave
expression~\cite{Schnedermann:1993ws},
\begin{equation}
\label{bw}
\left.\frac{\dd N_{\Psi}}{p_t\dd p_t}\right\vert_{coal}\propto 
m_t\int^R_0 rdr K_1\left(\frac{m_t\cosh y_t}{T}\right)
I_0\left(\frac{p_t\sinh y_t}{T}\right) \ 
\end{equation}
with $m_t$=$\sqrt{m_\Psi^2+p^2_t}$, transverse rapidity
$y_t$=$\tanh^{-1}v_t(r)$ with linear flow profile
$v_t(r)$=$v_{s}\frac{r}{R}$ and surface velocity $v_s$ given by the
fireball evolution formula, Eq.(\ref{Vfb}). We evaluate the blastwave
expression at the hadronization transition ($T_c$=180~MeV) and neglect
rescattering of $\Psi$'s in the hadronic phase.

\section{Centrality and Rapidity Dependence}
\label{num}
\begin{figure}[!t]
\centering
\includegraphics[width=0.40\textwidth]{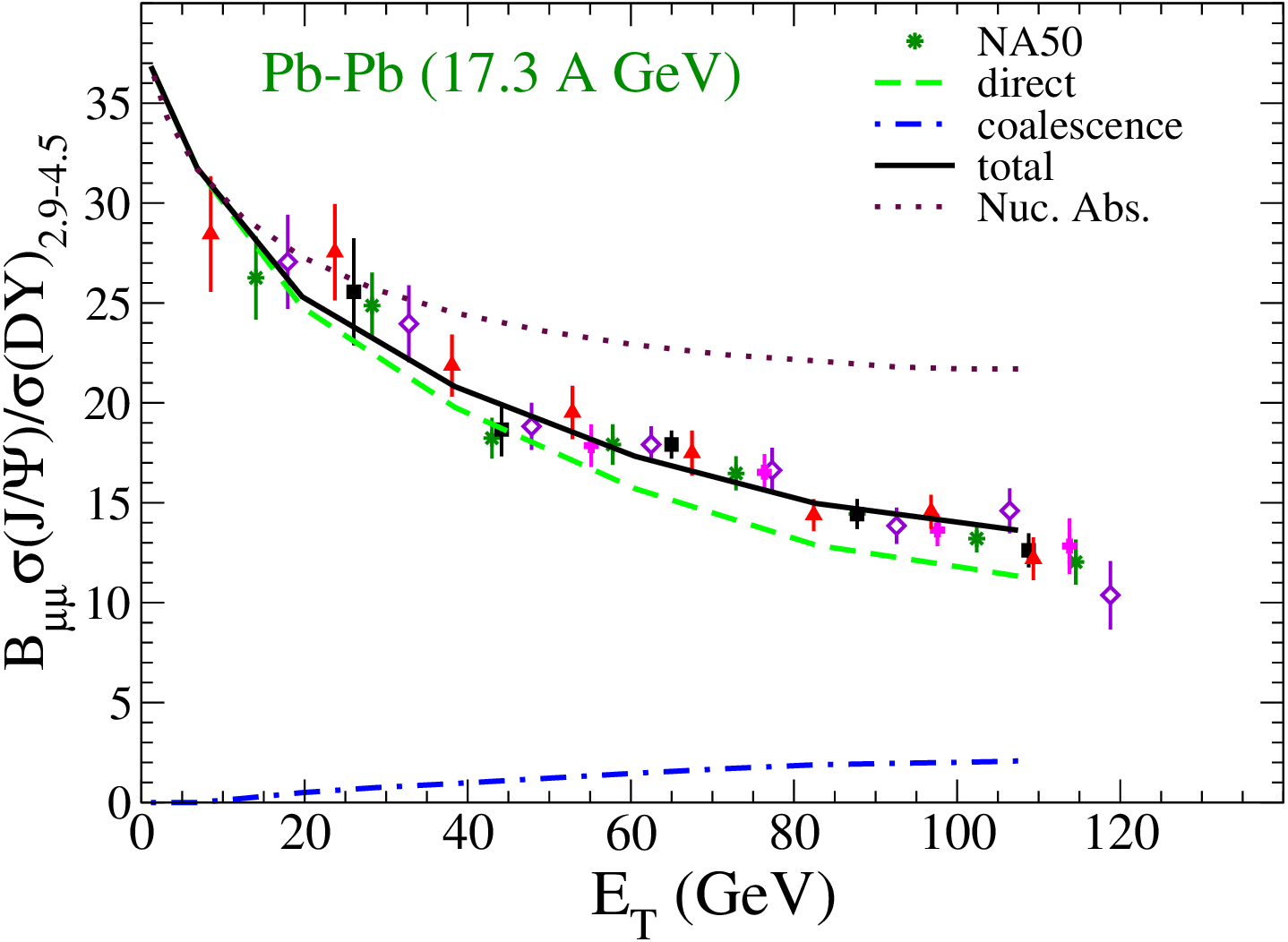}

\vspace{0.4cm}

\includegraphics[width=0.40\textwidth]{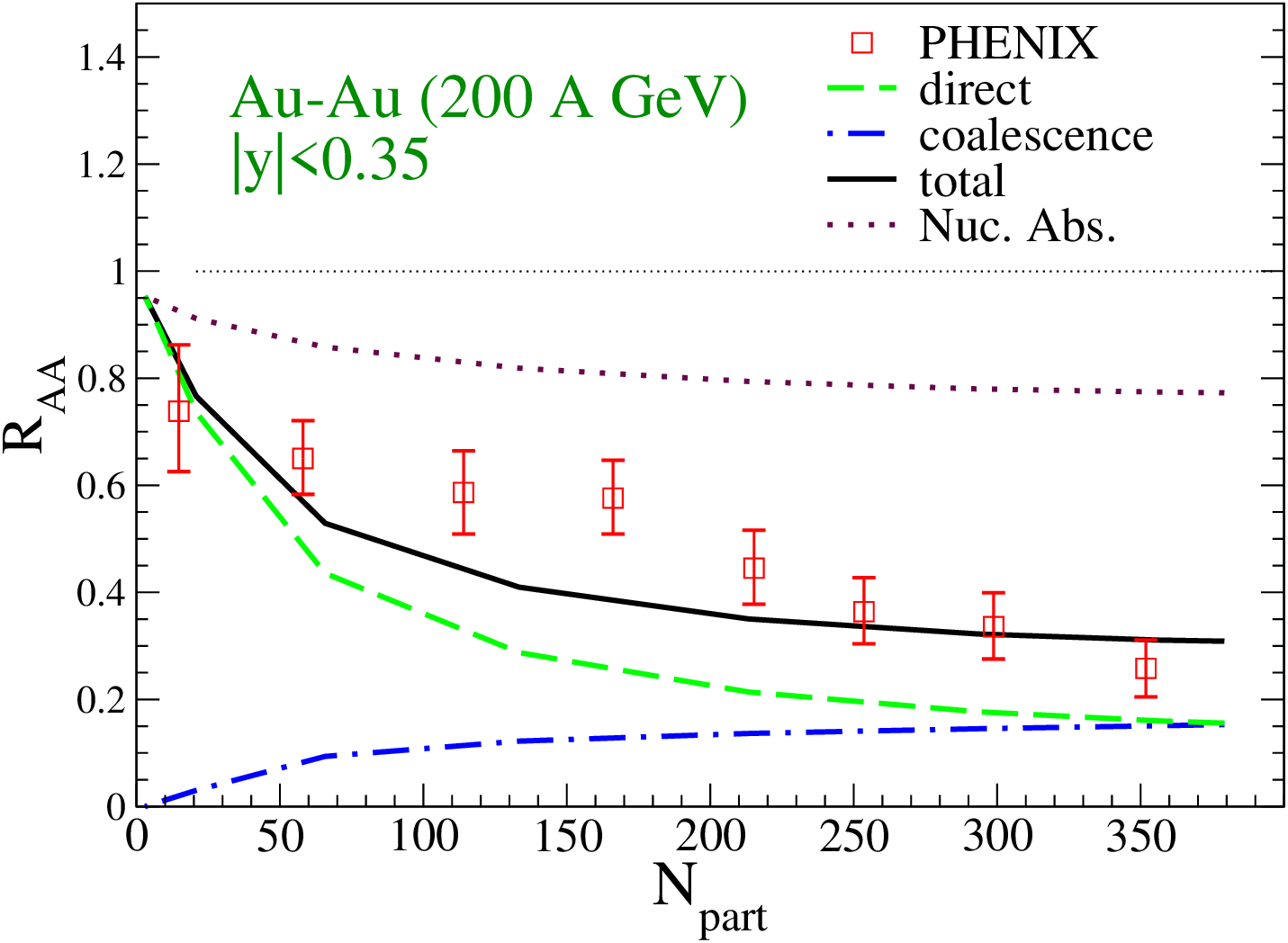}
\caption{(Color online) Results of the thermal rate-equation approach
  for $R_{AA}^{J/\psi}$ vs. centrality at SPS (upper panel) and RHIC
  midrapidity (lower panel) are compared to NA50~\cite{Ramello:2003ig}
  and PHENIX data~\cite{Adare:2006ns}. Solid line: total $J/\psi$
  yield; dashed line: suppressed primordial production; dot-dashed
  line: regeneration component; dotted line: primordial production
  with nuclear absorption only.}
\label{raa_centra_mid}
\end{figure}

The number of $J/\psi$ produced in initial hard collisions at given
impact parameter $(b)$ is estimated from the production cross section
in $p$-$p$ collisions and then scaled by number of binary collisions
in $N$-$N$ collisions, $N_{coll}(b)$. We use the $J/\psi$ production
cross section from PHENIX data~\cite{Adare:2006kf} with $\dd
\sigma_{pp}^{J/\psi}/\dd y$=750(500)~nb (with about 30\% uncertainty)
at mid (forward) rapidity. The centrality dependence of the inclusive
$J/\psi$ yield in terms of nuclear modification factor,
\begin{equation}
\label{raa}
R_{AA}=\frac {N_{J/\psi}^{AA}}{N_{J/\psi}^{pp}N_{coll}} \ ,
\end{equation}
is displayed in Fig.~\ref{raa_centra_mid}.
For central collisions the direct and regeneration component are about
equal (quite similar to Ref.~\cite{Yan:2006ve}), while for peripheral
collisions the direct component dominates because the lifetime of the
medium is shorter so that there is less time for regeneration
processes and for charm quarks to thermally equilibrate (reducing
coalescence into charmonia). Fig.~\ref{raa_centra_mid} also includes
the strength of suppression caused by CNM effects (dotted line) and by
the hot medium indicating the latter to be substantially larger than
the former.
\begin{figure}[!t]
\centering
\includegraphics[width=0.40\textwidth]{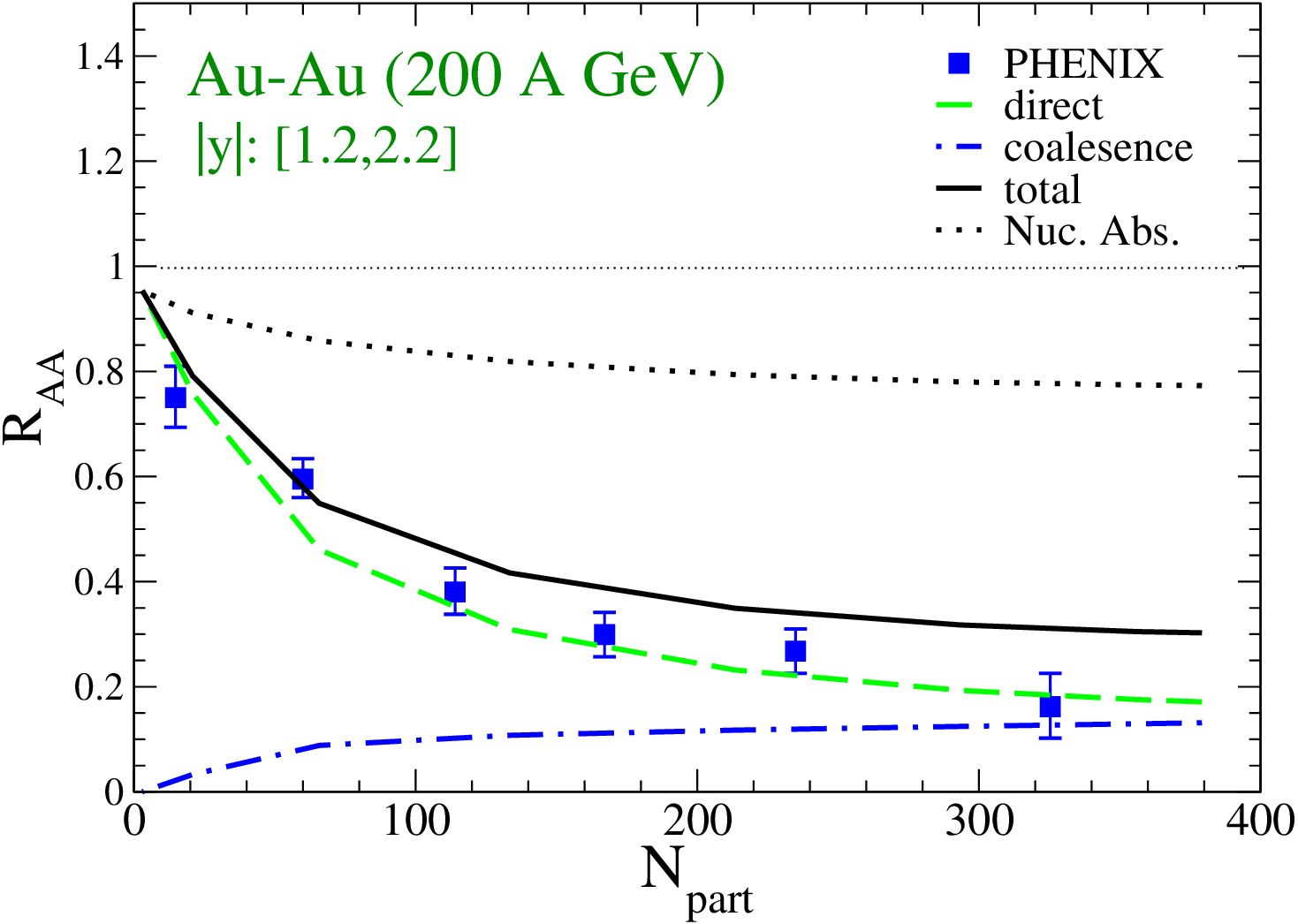}

\vspace{0.4cm}

\includegraphics[width=0.40\textwidth]{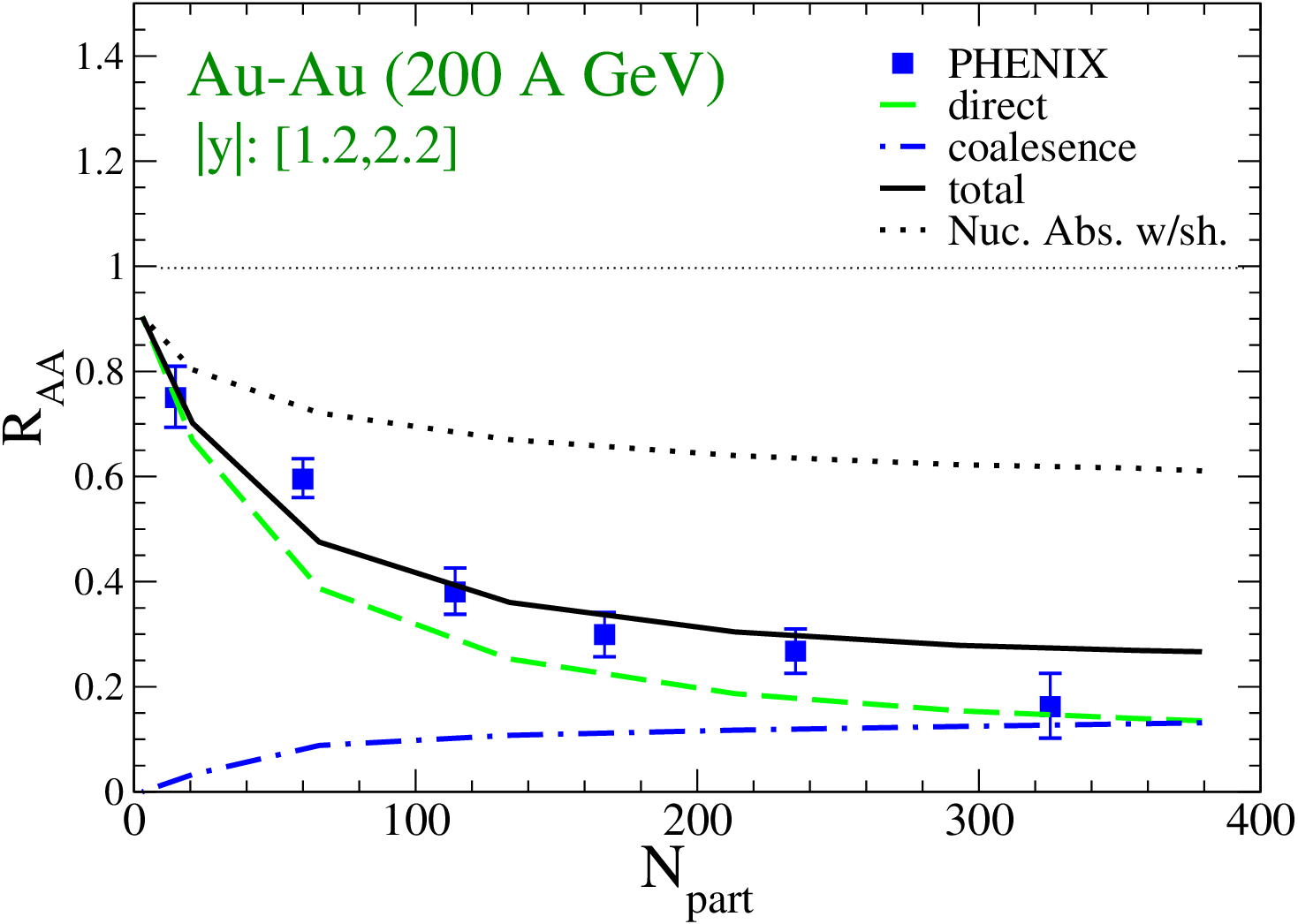}
\caption{(Color online) Results of the thermal rate-equation approach
  for $R_{AA}^{J/\psi}$ vs. centrality at forward rapidity compared to
  PHENIX data~\cite{Adare:2006ns} with CNM effects in scenario 1
  (upper panel) and scenario 2 (lower panel). Solid line: total
  $J/\psi$ yield; dashed line: suppressed primordial production;
  dot-dashed line: regeneration component; dotted line: primordial
  production with CNM effects only. }
\label{raa_centra_forw}
\end{figure}
\begin{figure}[!t]
\centering
\includegraphics[width=0.40\textwidth]{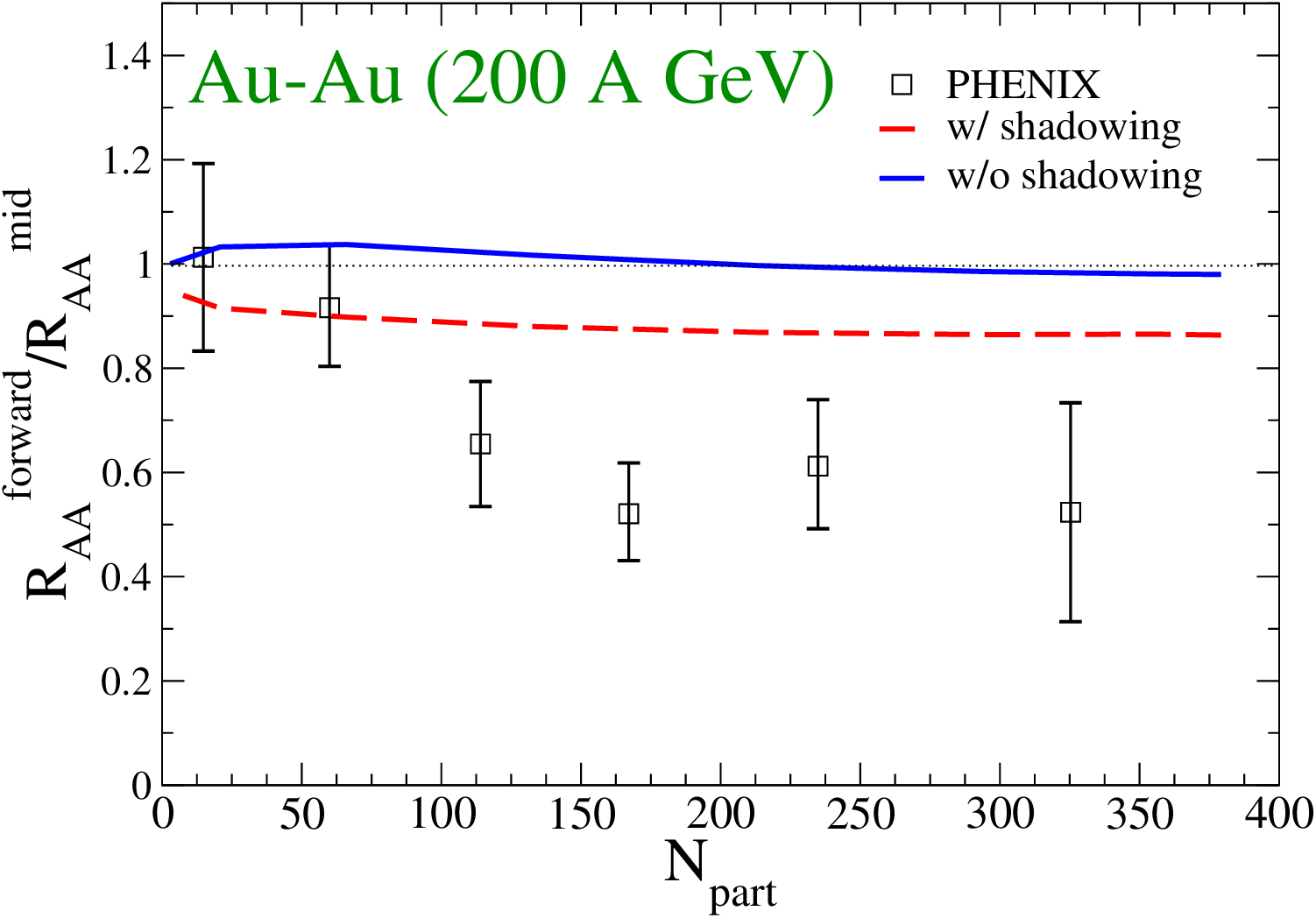}
\caption{(Color online) The thermal rate-equation approach results of
  ratio of forward and midrapidity $J/\psi$ $R_{AA}$ versus $N_{part}$
  with cold nuclear matter effects in scenario 1 (solid line) and
  scenario 2 (dashed line) compared to PHENIX
  data~\cite{Adare:2006ns}.}
\label{raa_centra_forw_mid}
\end{figure}

Next we proceed to the inclusive $J/\psi$ yield at forward rapidity as
shown in Fig.~\ref{raa_centra_forw} with scenarios 1 and 2 for CNM
effects. In scenario 1, where the CNM induced suppression at forward
rapidity is assumed to be the same as at mid-rapidity, the hot medium
causes slightly less suppression than at mid-rapidity due to a shorter
QGP lifetime (recall Fig.~\ref{fig_fb}). Concerning the regeneration
component, we note that the number of charm quark pairs at RHIC energy
is between the canonical and grandcanonical limit, so that
$N_{\Psi}^{stat}\sim N_{c\bar c}^\alpha$ with $\alpha$ between 1
(canonical limit) and 2 (grandcanonical limit) as following from the
charm conservation equation (\ref{eq:Ncc}). If charm and $J/\psi$
production in $p$-$p$ collisions (the denominator in $R_{AA}$) are
less at forward rapidity than at midrapidity by about the same
fraction, the reduction of regeneration component
at forward rapidity is about $R_{AA}\sim N_{c\bar
  c}^{\alpha-1}$. Adding up the two components the total $J/\psi$
yield at forward rapidity is almost equal to that at midrapidity since
the slight decrease in the regeneration component is compensated by
the slight increase in the direct component, see upper panel of
Fig.~\ref{raa_centra_forw}. Scenario 2 differs from scenario 1 in that
the direct component at forward rapidity is subject to stronger CNM
suppression. Consequently, the inclusive $J/\psi$ yield at forward
rapidity is more suppressed which leads to better agreement with
experimental data as seen in the lower panel of
Fig.~\ref{raa_centra_forw}. To better illustrate the comparison
between forward and midrapidity we plot the ratio of forward and
midrapidity $J/\psi$ $R_{AA}$ and compare to experimental data in
Fig.~\ref{raa_centra_forw_mid}. Even with a stronger cold matter
induced suppression at forward rapidity due to shadowing (scenario~2)
our approach still appears not to fully account for the experimental
findings. However one needs to keep in mind that the current
uncertainties in several key inputs, in particular charm production
cross section at forward rapidity and shadowing, are still
appreciable. More accurate $d$-$Au$ data will enable more definite
conclusions.
\begin{figure}[!t]
\centering
\includegraphics[width=0.40\textwidth]{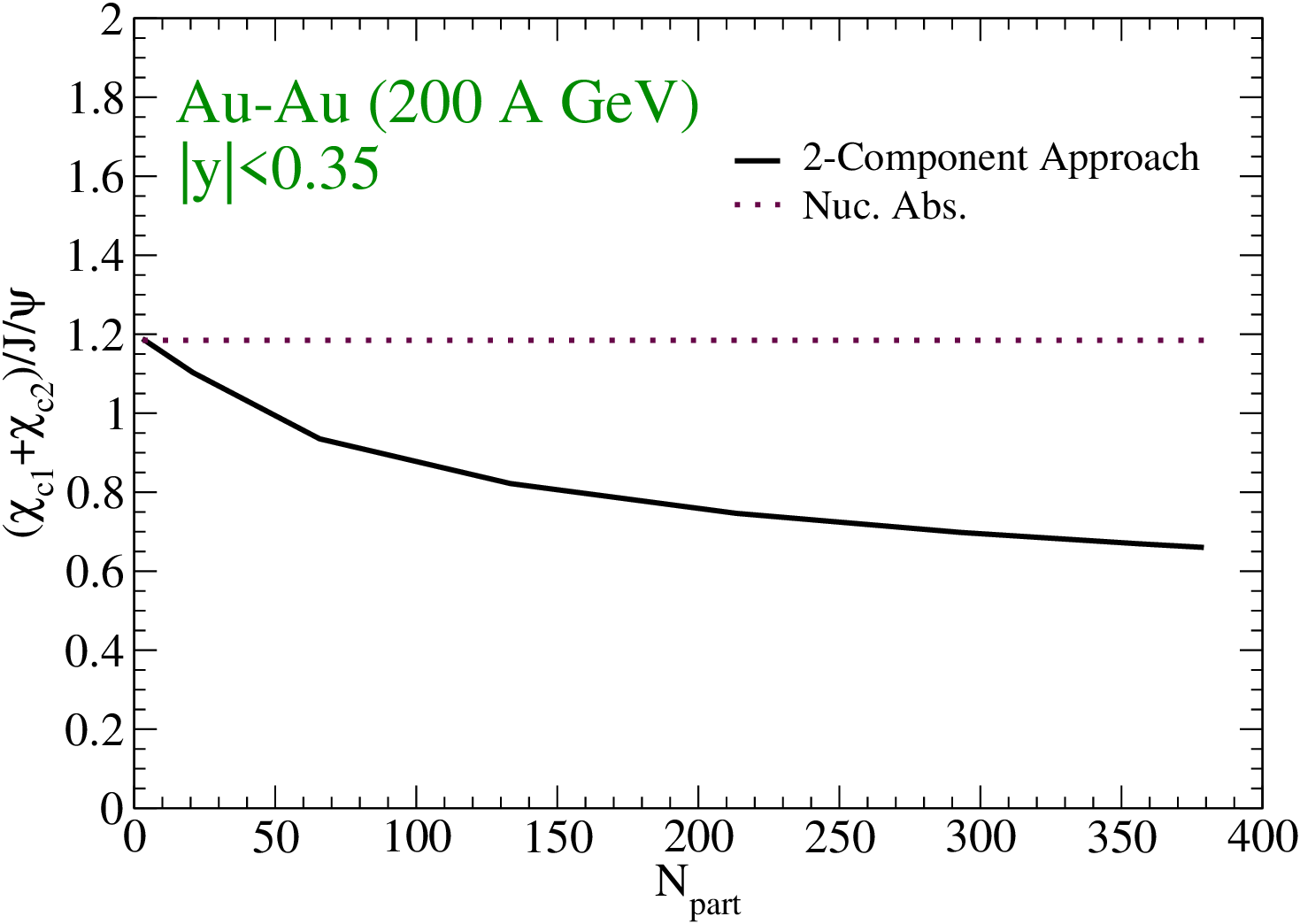}

\vspace{0.4cm}

\includegraphics[width=0.40\textwidth]{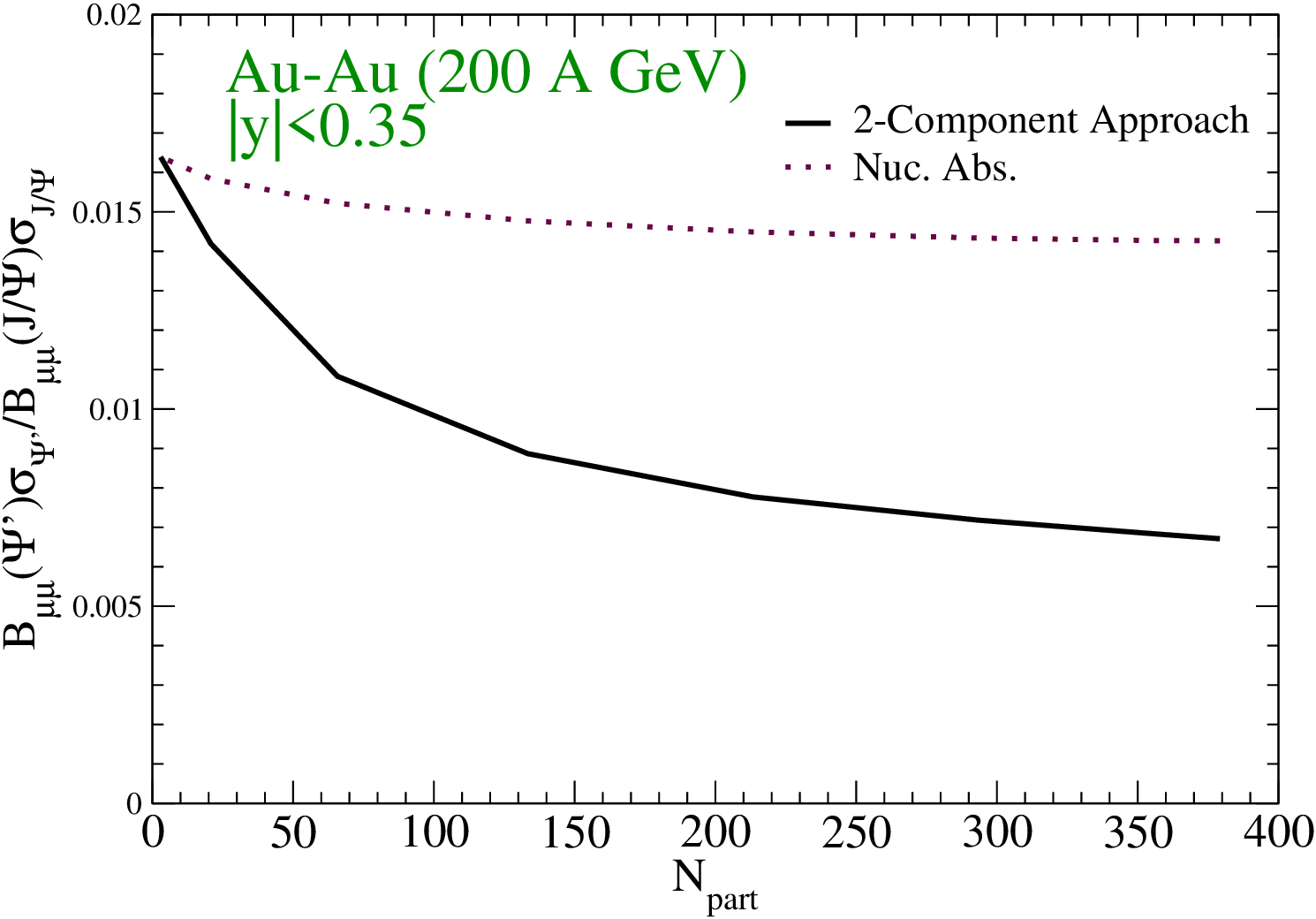}
\caption{(Color online) Results of the thermal rate-equation approach
  for the $(\chi_{c1}+\chi_{c2})/(J/\psi)$ ratio (upper panel) and
  $\psi'/(J/\psi)$ ratio (lower panel) vs. centrality at midrapidity
  at RHIC. Solid line: the results of the thermal rate-equation
  approach; dashed line: the nuclear absorption only. The $J/\psi$'s
  in the denominator include feeddown from $\chi_c$ and $\psi'$.}
\label{chi_c_psi}
\end{figure}

Finally we calculate the $\chi_c/(J/\psi)$ and $\psi'/(J/\psi)$ ratios
which may provide additional discrimination of charmonium production
mechanisms: \eg, the gluo-dissociation process and quasifree
dissociation process give comparable suppression for $J/\psi$ but for
$\chi_c$ the gluo-dissociation process gives much larger dissociation
rate~\cite{Zhao:2007hh} leading to a much smaller $\chi_c/(J/\psi)$
ratio. As another example, if formation time
effects~\cite{Blaizot:1988ec,Gavin:1990gm,Karsch:1988} are important
one may observe less suppression of $\chi_c$ than $J/\psi$ due to a
longer~\cite{Karsch:1988} formation time of $\chi_c$ compared to
$J/\psi$, together with a smaller dissociation cross section for a
``pre-hadronic'' $c\bar c$ pair than for a fully formed
charmonium. This is in contrast to most standard dissociation
mechanisms (since $\chi_c$ is a higher excited $c\bar c$ state than
$J/\psi$ with smaller binding energy and therefore is more easily
destroyed) and regeneration mechanism ($\chi_c$ is heavier so that its
equilibrium abundance is suppressed compared to $J/\psi$ by the
Boltzmann thermal factor).

For $\chi_c$ states, we constrain ourselves to $\chi_{c1}$ and
$\chi_{c2}$ with a combined average branching ratio of 27\% into
$J/\psi$'s \cite{Yao:2006px}. The resulting $\chi_c/(J/\psi)$ and $\psi'/(J/\psi)$ ratio
at midrapidity are shown in Fig.~\ref{chi_c_psi}. Both drop with
centrality below the ratios obtained from CNM-induced suppression (we
assume that $J/\psi$ and $\chi_c$ undergo the same strength of
CNM-induced suppression).

\section{Transverse Momentum Spectra}
\label{pt}
\begin{figure}[!t]
\centering
\includegraphics[width=0.40\textwidth]{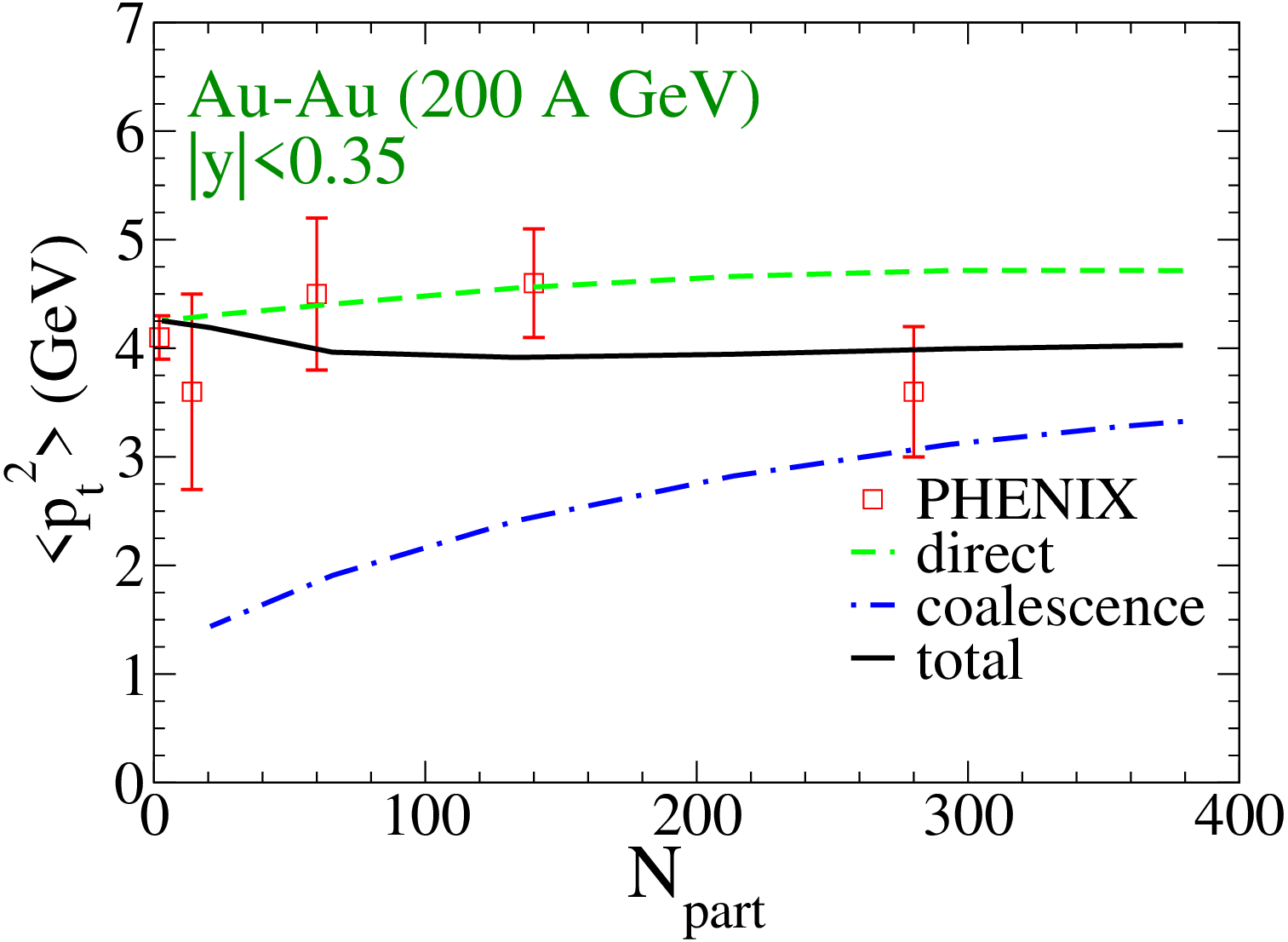}

\vspace{0.4cm}

\includegraphics[width=0.40\textwidth]{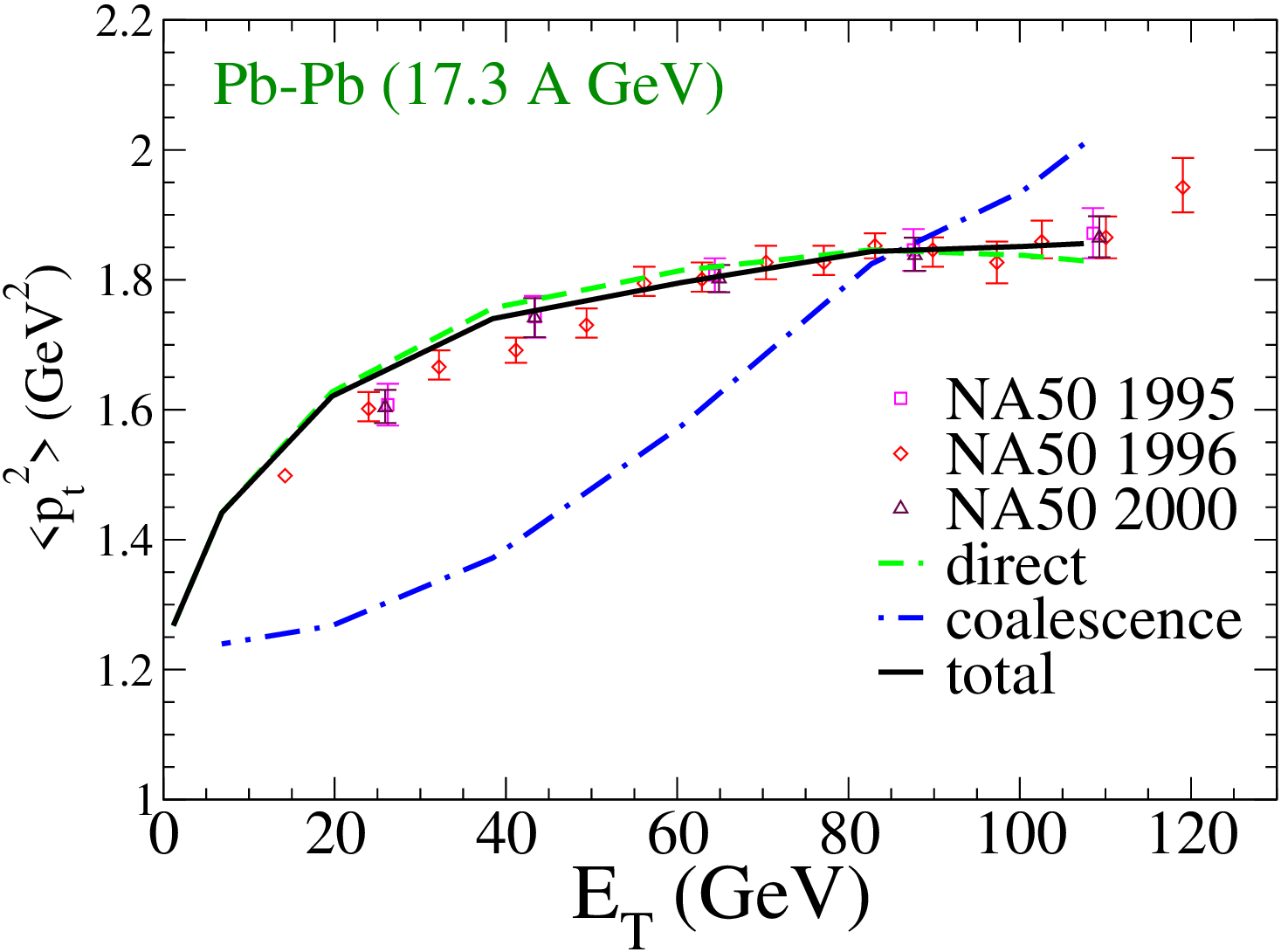}
\caption{(Color online) Results of the thermal rate-equation approach
  for $\langle p^2_t\rangle$ vs. centrality at RHIC midrapidity (upper
  panel) and SPS (lower panel) compared to PHENIX~\cite{Adare:2006ns}
  and NA50 data~\cite{Abreu:2000xe,Topilskaya:2003iy}. Solid line:
  $\langle p^2_t\rangle$ for the total $J/\psi$ yield (direct
  component + regeneration component); dashed line: $\langle
  p^2_t\rangle$ for the direct component; dot-dashed line: $\langle
  p^2_t\rangle$ for the regeneration component.}
\label{pt2_centra_mid}
\end{figure}

In order to gain further insights into charmonium production
mechanisms, and to discriminate among them, we investigate the transverse momentum spectra of
$J/\psi$. Let us again start at midrapidity: The centrality
dependence of $\langle p^2_t\rangle$ is shown in the upper panel of
Fig.~\ref{pt2_centra_mid}. The $\langle p^2_t\rangle$ for the direct
component increases with centrality due to a stronger Cronin effect
generated by a longer path length $\langle l\rangle$ travelled by the
gluon pair (recall Section~\ref{initial}) in more central collisions.
The $\langle p^2_t\rangle$ of the regeneration component also
increases with centrality due to an increase of the radial flow
velocity at the end of mixed phase. However, the magnitude of $\langle
p^2_t\rangle$ of the regeneration component is much smaller than
primordial production at all centralities. The main point is that for
more central collisions the regeneration component makes up an
increasing fraction of the total yield leading to an almost constant
$\langle p^2_t\rangle$ with centrality, consistent with current PHENIX
data. Similar results are also obtained by the kinetic recombination
model of Ref.~\cite{Thews:2005vj}. On the contrary, at SPS energy, the direct
component dominates at all centrality so that the Cronin effect
dominates leading to a monotonically increasing $\langle
p^2_t\rangle$, see lower panel of Fig.~\ref{pt2_centra_mid}. A decreasing $\langle p^2_t\rangle$ with centrality is
thus a supporting signature of the presence of regeneration. More
definite conclusions will be possible with improved data accuracy.
\begin{figure}[!t]
\centering
\includegraphics[width=0.40\textwidth]{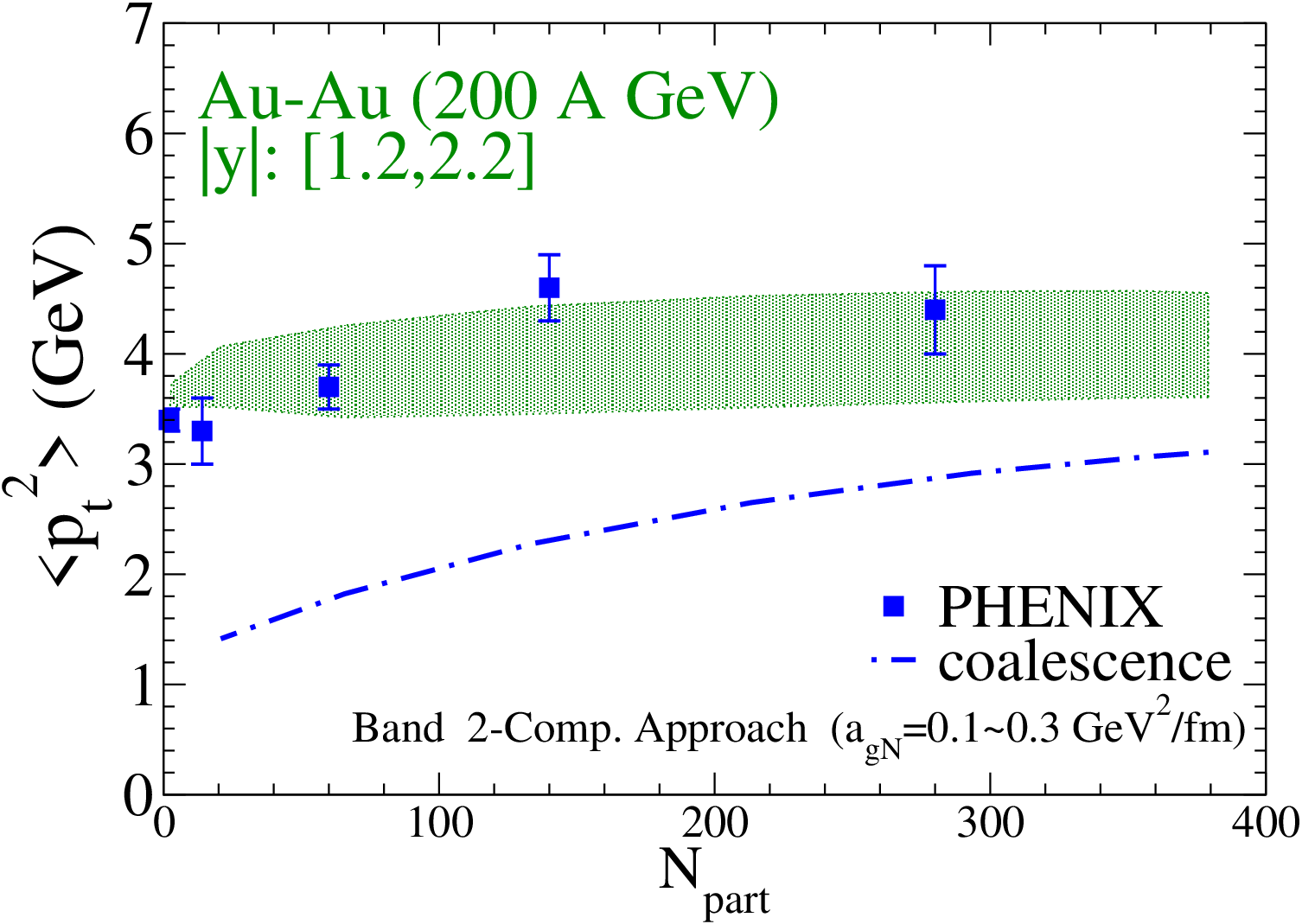}
\caption{(Color online) Results of the thermal rate-equation approach
  for $\langle p^2_t\rangle$ vs. centrality at forward rapidity
  compared to PHENIX data~\cite{Adare:2006ns}. The band represents the
  $\langle p^2_t\rangle$ for the total $J/\psi$ yield (direct
  + regeneration). The width of the band
  represents the uncertainty from the strength of the Cronin effect
  (estimated from $d$-$Au$ data with $a_{gN}$=0.1-0.3~GeV$^2$/fm). The
  cold nuclear matter effects are implemented in scenario 1. Dot-dashed
  line: $\langle p^2_t\rangle$ for the regeneration component.}
\label{pt2_centra_forw}
\end{figure}

We now turn to the $\langle p^2_t\rangle$ vs. centrality at forward
rapidity as shown in Fig.~\ref{pt2_centra_forw} (with CNM effects in
scenario 1). Within the current uncertainties in the Cronin effect
our results agree with experimental data. The increase of $\langle p^2_t\rangle$ suggests that
the direct component at forward rapidity is substantial; otherwise the
$\langle p^2_t\rangle$ would be significantly smaller.
\begin{figure*}[!t]
\includegraphics[width=0.40\textwidth]{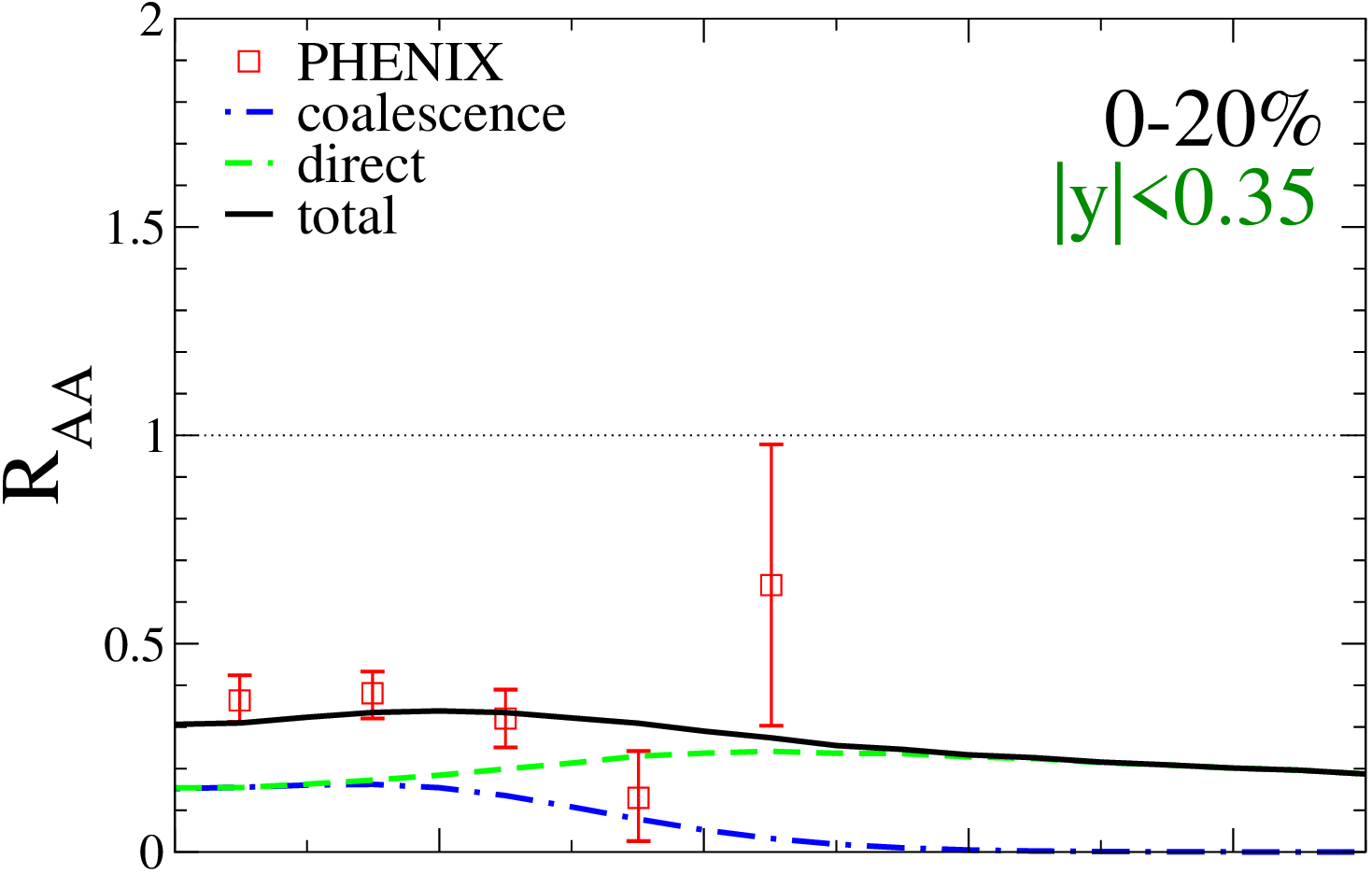}
\hfill
\includegraphics[width=0.40\textwidth]{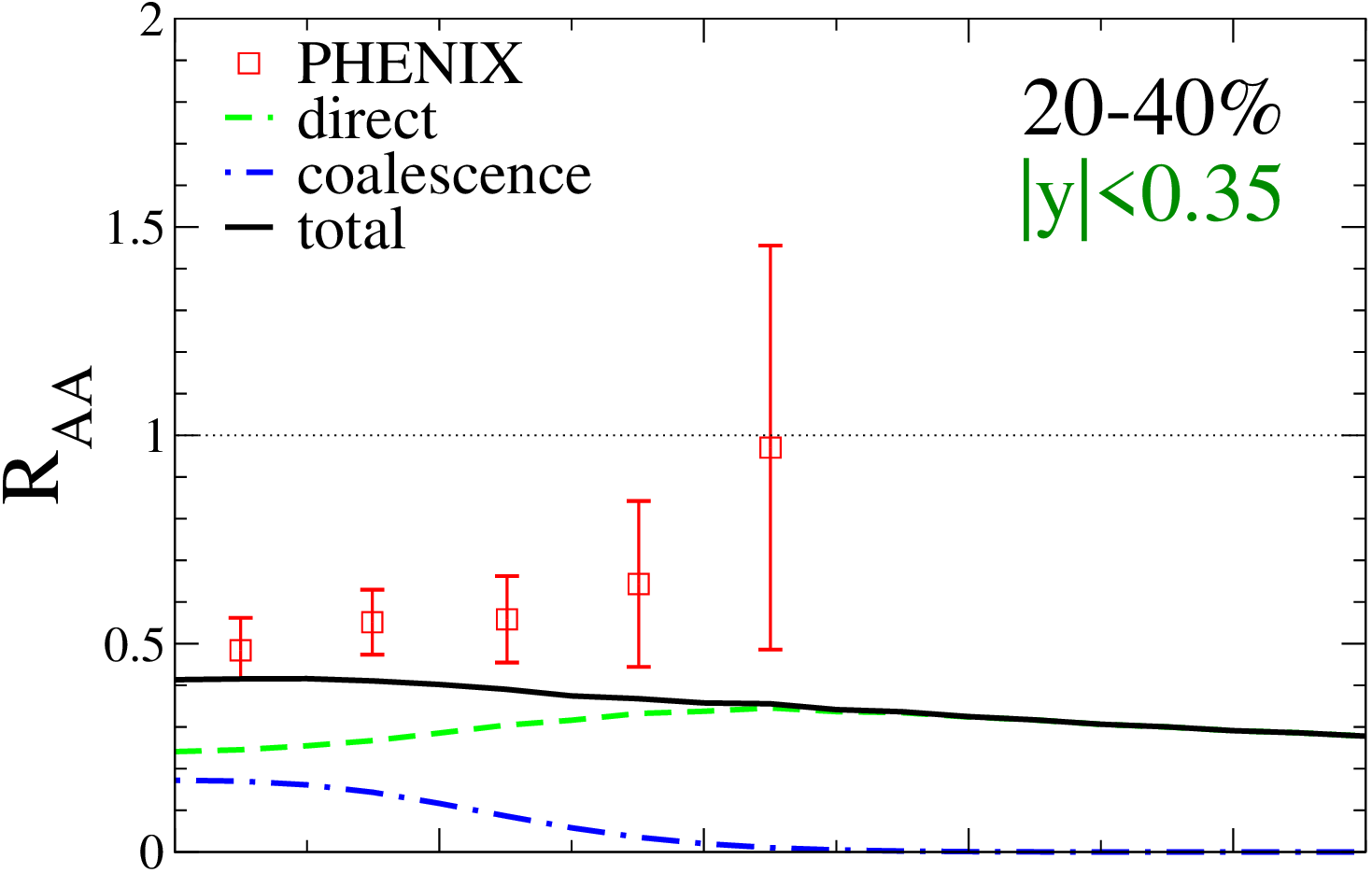}
\includegraphics[width=0.40\textwidth]{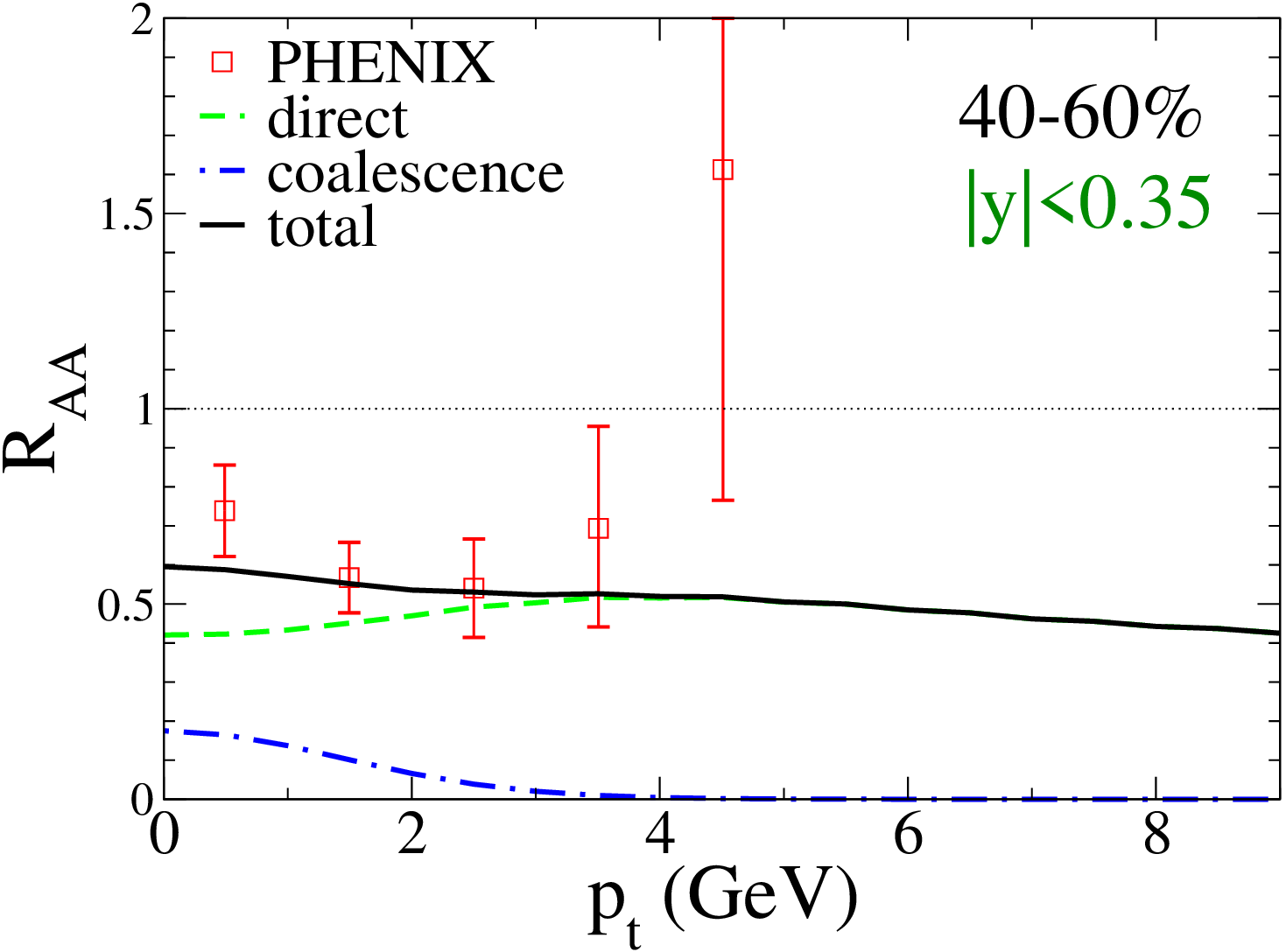}
\hfill
\includegraphics[width=0.40\textwidth]{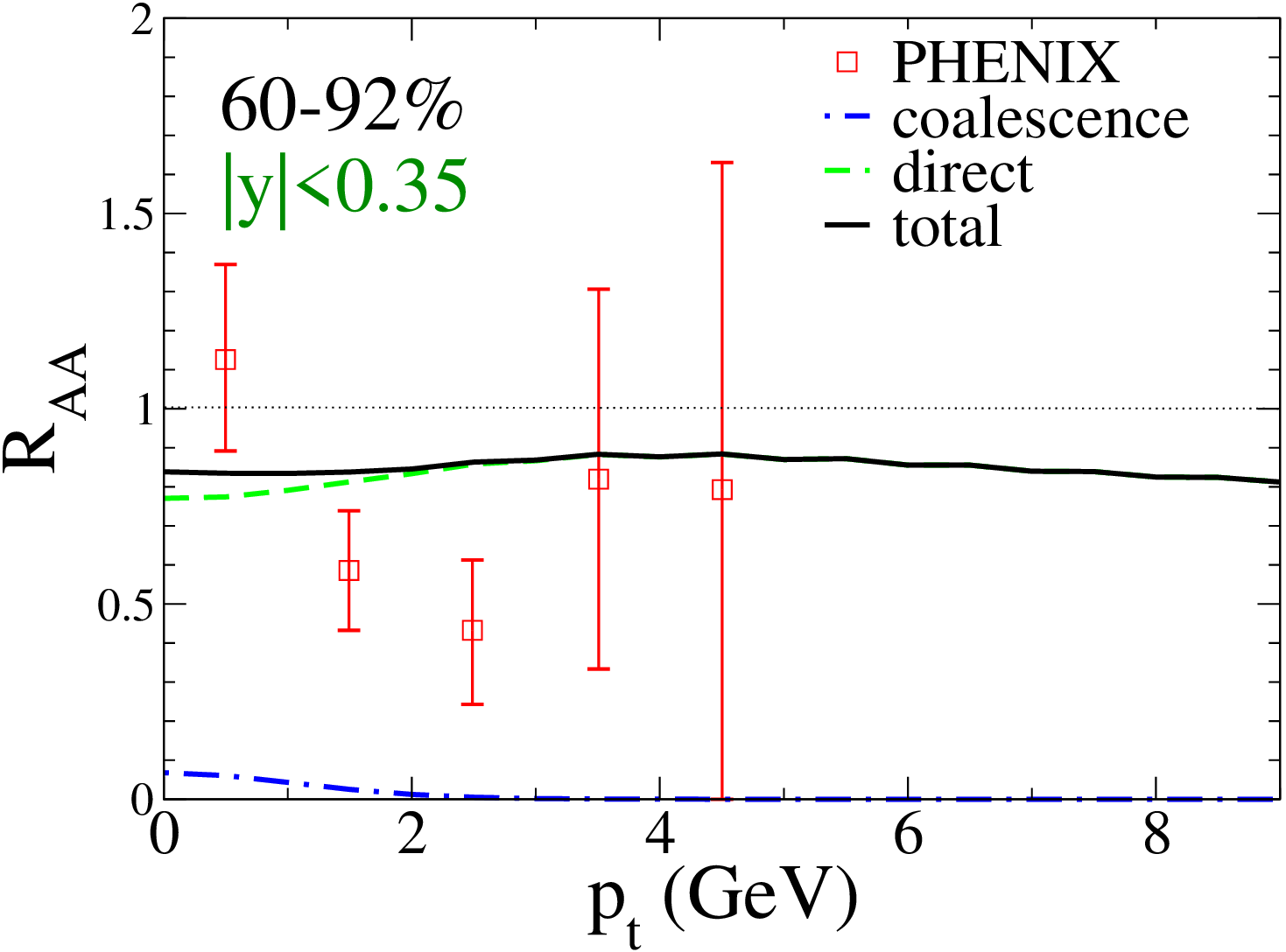}
\caption{(Color online) Results of the thermal rate-equation approach
  for $R_{AA}^{J/\psi}$ vs. transverse momentum for different
  centrality selections at midrapidity are compared to PHENIX
  data~\cite{Adare:2006ns}. Solid line: total $J/\psi$ yield; dashed
  line: suppressed primordial production; dot-dashed line:
  regeneration component. The Cronin effects are implemented with $a_{gN}$=0.1~GeV$^2$/fm.}
\label{raa_pt_mid}
\end{figure*}
\begin{figure*}[!t]
\includegraphics[width=0.40\textwidth]{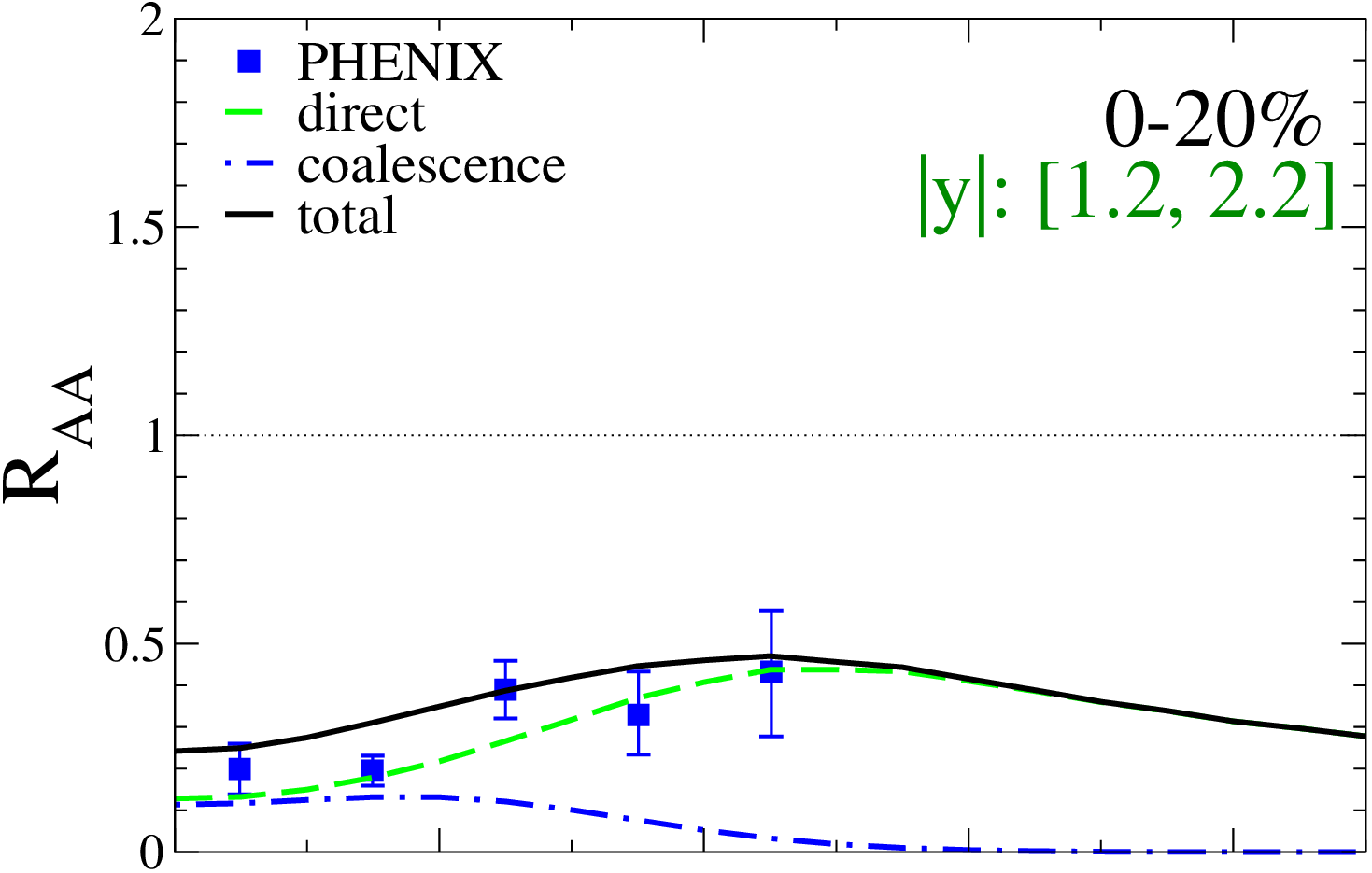}
\hfill
\includegraphics[width=0.40\textwidth]{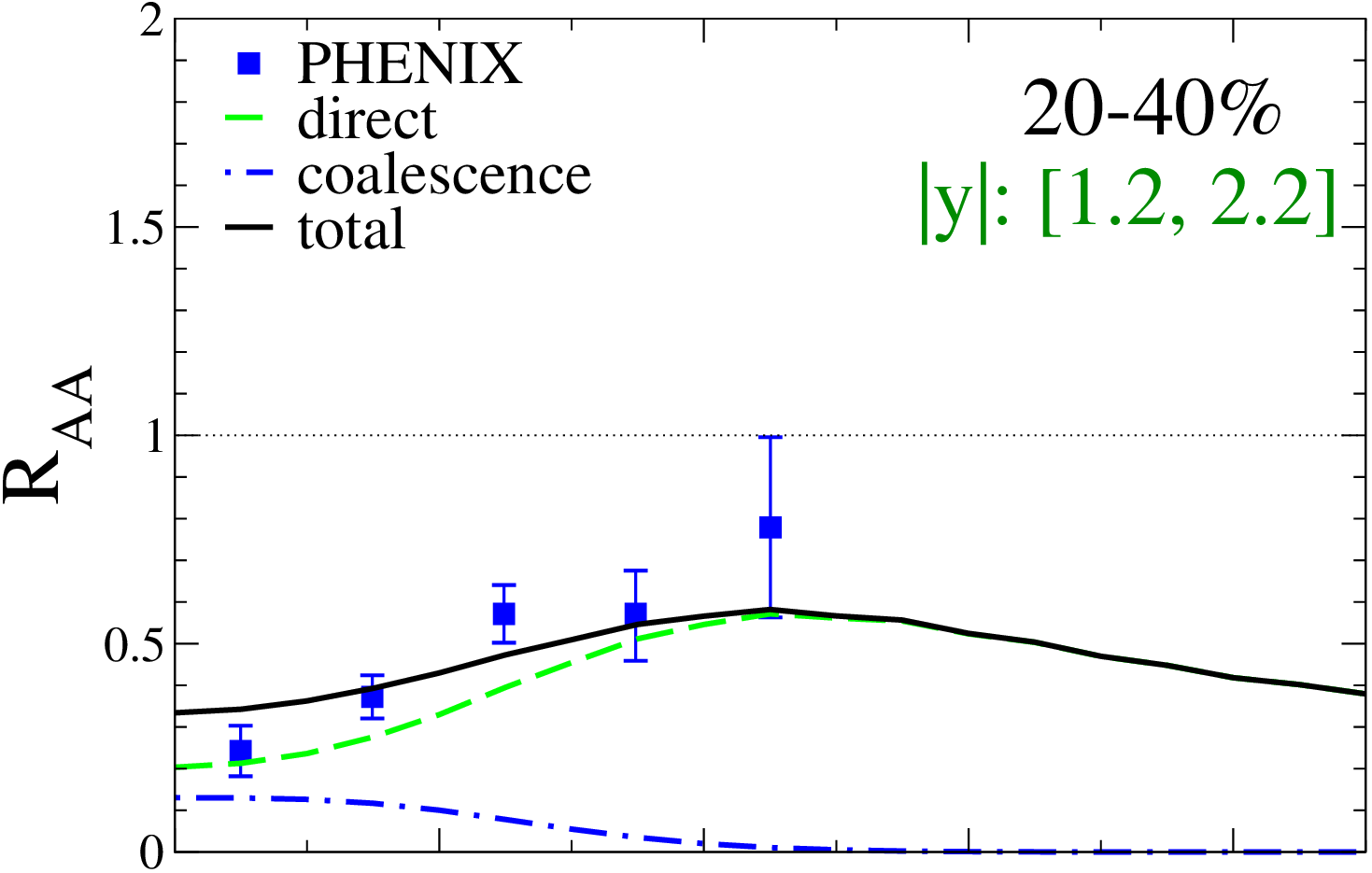}
\includegraphics[width=0.40\textwidth]{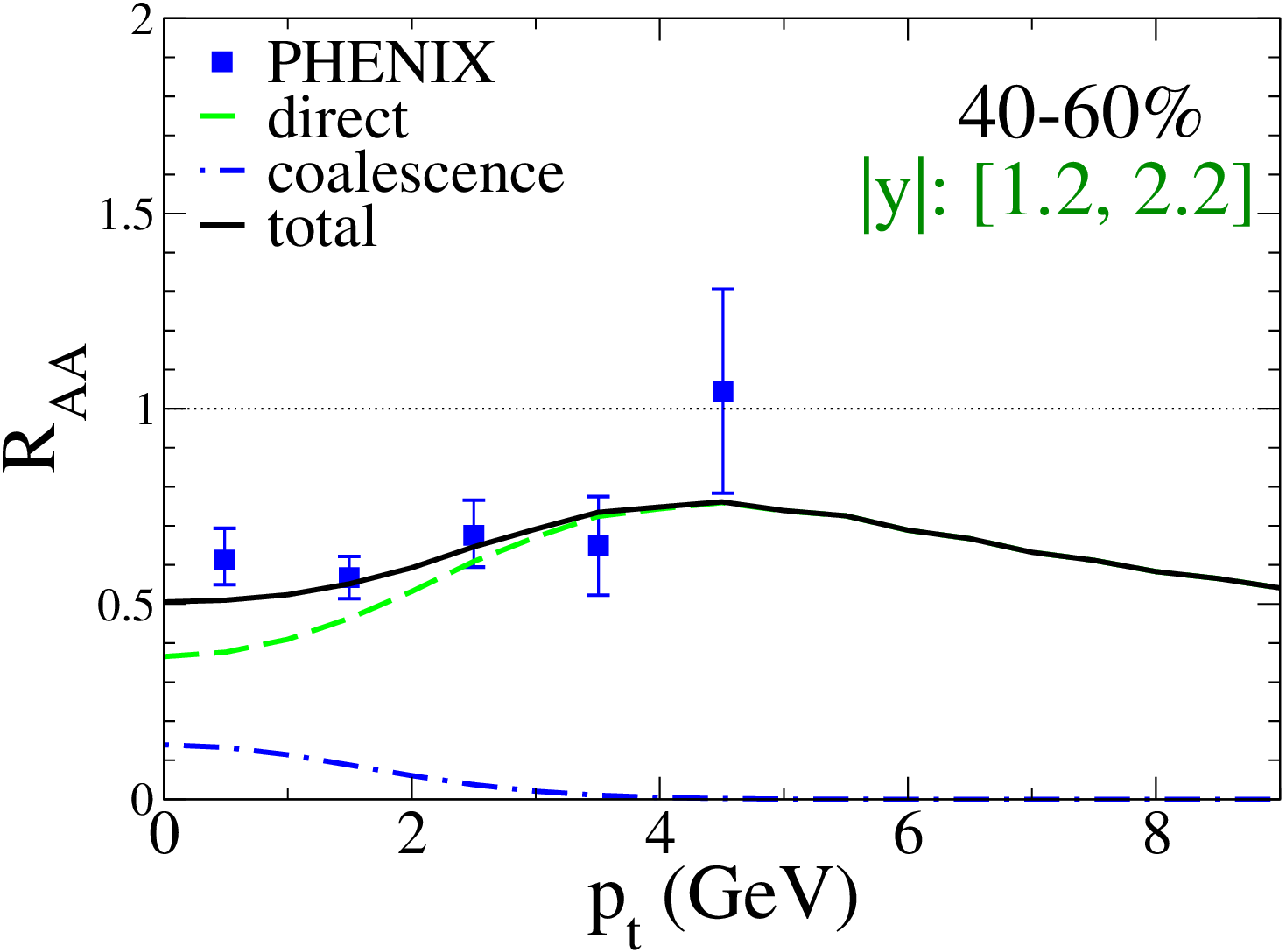}
\hfill
\includegraphics[width=0.40\textwidth]{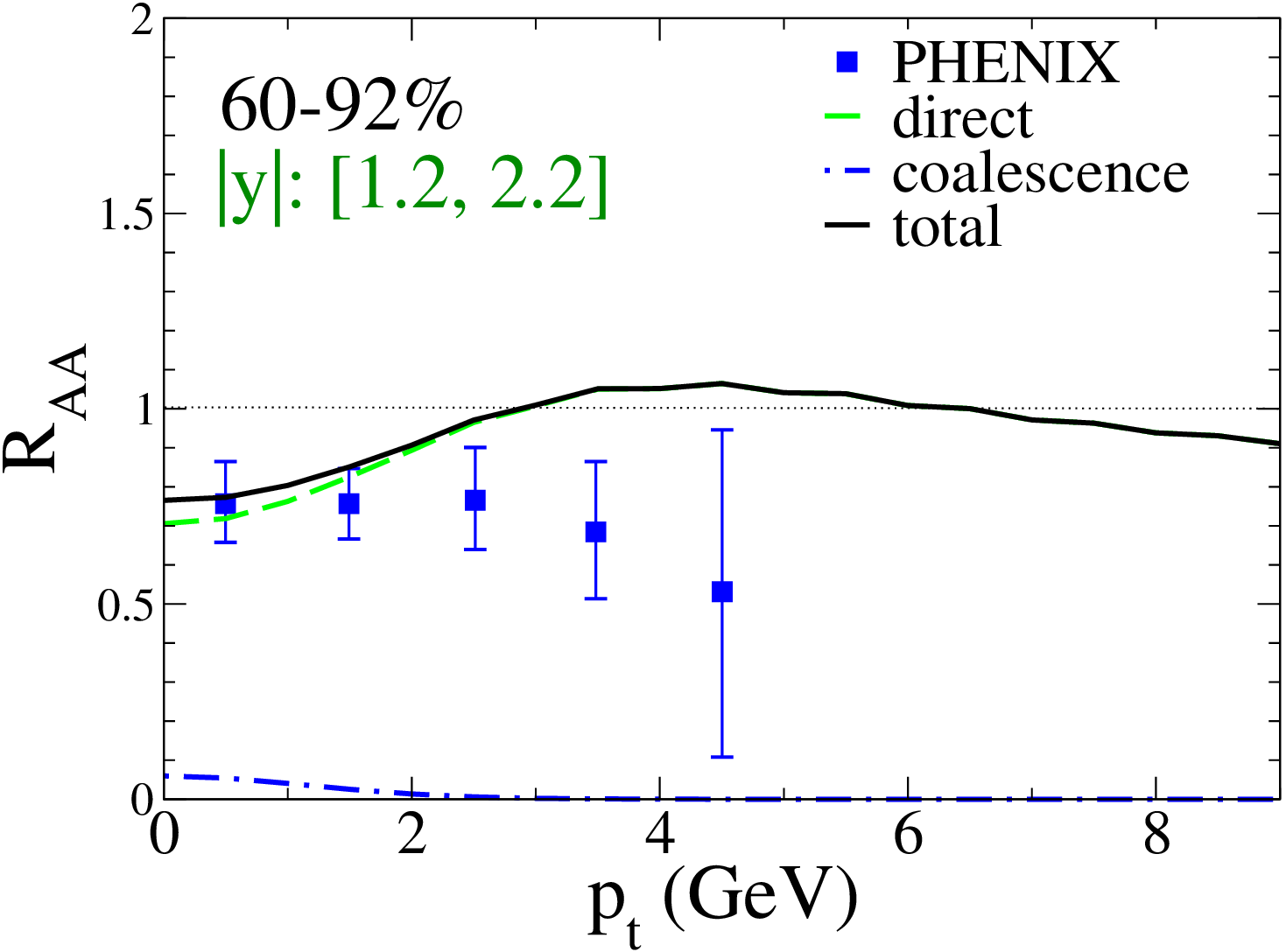}
\caption{(Color online) Results of the thermal rate-equation approach
  for $R_{AA}^{J/\psi}$ vs. transverse momentum for different
  centrality selections at forward rapidity are compared to PHENIX
  data~\cite{Adare:2006ns}. Solid line: total $J/\psi$ yield; dashed
  line: suppressed primordial production; dot-dashed line:
  regeneration component. The CNM effects are implemented in scenario
  1 with $a_{gN}$=0.2~GeV$^2$/fm.}
\label{raa_pt_forw}
\end{figure*}

For a more differential comparison with experimental data we plot
$R_{AA}(p_t)$ for different centrality selections at both
mid- and forward rapidity (with CNM effects in
scenario 1 and $a_{gN}$=0.2~GeV$^2$/fm) in Fig.~\ref{raa_pt_mid} and~\ref{raa_pt_forw}
respectively. Overall, our model results reproduce $R_{AA}(p_t)$ data reasonably
well. By comparing forward and midrapidity results there is an indication for 
stronger suppression at forward rapidity relative to midrapidity
mainly in the low $p_t$ regime (for central and semicentral
collisions). Less regeneration and stronger Cronin effect at forward rapidity can both generate such an effect.

\section{Conclusion}
\label{concl}
In this work we have applied a previously constructed thermal
rate-equation approach to study charmonium production in 200 AGeV
$Au$-$Au$ collisions at RHIC at forward rapidity and compared to
midrapidity results. Our calculations reasonably reproduce the
experimental data of inclusive $J/\psi$ $R_{AA}$, $\langle
p^2_t\rangle$ and $p_t$ spectra at forward rapidity. We find that the
observed stronger suppression at forward rapidity can be partially
attributed to smaller regeneration resulting from smaller open charm
production but additional suppression from cold nuclear matter effects
(shadowing) is required as well. An accurate knowledge of shadowing
will be needed to clarify the ``puzzle'' of the stronger suppression
at forward rapidity. As a next step we will improve our current
treatment of the coalescence of $c$ and $\bar c$ quarks by performing
a microscopic calculation with a time-dependent charm quark
distribution based on Ref.~\cite{vanHees:2007me}, which will allow for
a more accurate evaluation of regeneration effects, especially their
dependence on incomplete charm-quark equilibration. We also look
forward to new experimental data to better constrain the inputs of our
approach which will pave the way for more quantitative conclusions.

%
%
%
%
%

\end{document}